\documentclass[sigconf, nonacm]{acmart}

\usepackage{placeins}
\usepackage{graphicx}
\usepackage{pifont}
\usepackage{subcaption}
\usepackage{algorithm}
\usepackage[noend]{algpseudocode}
\algrenewcommand\algorithmicrequire{\textbf{Input:}}
\algrenewcommand\algorithmicensure{\textbf{Output:}}
\algrenewcommand\algorithmicindent{1em} 
\usepackage{multirow}

\usepackage{amsmath}
\usepackage{amsthm}
\theoremstyle{plain}

\theoremstyle{definition}
\newtheorem{definition}{Definition}
\theoremstyle{remark}

\newcommand\vldbdoi{XX.XX/XXX.XX}
\newcommand\vldbpages{XXX-XXX}
\newcommand\vldbvolume{14}
\newcommand\vldbissue{1}
\newcommand\vldbyear{2020}
\newcommand\vldbauthors{\authors}
\newcommand\vldbtitle{\shorttitle} 
\newcommand\vldbavailabilityurl{https://github.com/QDXG-CXK/FROG}
\newcommand\vldbpagestyle{plain}

\begin{document}
\title{FROG: Efficient Range-Filtering Approximate Nearest Neighbor Search on GPUs}

\settopmatter{authorsperrow=4}

\author{Xiaokun Cui}
\affiliation{%
  \institution{Xidian University}
  \city{Xi'an}
  \country{China}
}
\email{cuixk@stu.xidian.edu.cn}

\author{Pengbo Liu}
\affiliation{%
  \institution{Xidian University}
  \city{Xi'an}
  \country{China}
}
\email{liupengbo@stu.xidian.edu.cn}

\author{Jiadong Xie}
\affiliation{%
  \institution{The Chinese University of Hong Kong}
  \city{Hong Kong SAR}
  \country{China}
}
\email{jdxie@se.cuhk.edu.hk}

\author{Yingfan Liu}
\authornote{Corresponding author.}
\affiliation{%
  \institution{Xidian University}
  \city{Xi'an}
  \country{China}
}
\email{liuyingfan@xidian.edu.cn}

\author{Hui Li}
\affiliation{%
  \institution{Xidian University}
  \city{Xi'an}
  \country{China}
}
\email{hli@xidian.edu.cn}

\author{Jeffrey Xu Yu}
\affiliation{%
  \institution{The Hong Kong University of Science and Technology (Guangzhou)}
  \city{Guangzhou}
  \country{China}
}
\email{jeffreyxuyu@hkust-gz.edu.cn}

\author{Jiangtao Cui}
\affiliation{%
  \institution{Xi'an University of Posts and Telecommunications}
  \city{Xi'an}
  \country{China}
}
\affiliation{%
  \institution{Xidian University}
  \city{Xi'an}
  \country{China}
}
\email{cuijt@xupt.edu.cn}

\renewcommand{\shortauthors}{Cui et al.}

\begin{abstract}
Range-filtering approximate nearest neighbor search (RFANNS) is a fundamental operation in modern vector databases. Given a query vector $q$ and a numerical range predicate, RFANNS returns the $k$-approximate nearest neighbors ($k$-ANN) of the query $q$ among the objects whose attributes satisfy the range predicate. However, existing RFANNS methods are not well suited to high-throughput GPU execution. CPU indexes offer limited parallel scalability, generic GPU filtering is highly selectivity-dependent, and GPU indexes built from locally optimized subgraphs can incur long search trajectories and redundant distance
computations.
To address these limitations, we present FROG, a GPU-oriented RFANNS index that replaces multiple locally optimal substructure building with a globally aware, vertex-centric design.  It organizes diverse expansion neighbor candidates for each vertex in a GPU-friendly structure and rapidly identifies the expansion neighbors used for computation at query time. Moreover, GPU-oriented algorithms and implementations are developed for both index construction and query processing.
Experiments on six datasets show that FROG improves mixed-selectivity
query throughput by 14.7--37.7$\times$ over 44-core CPU baselines and
4.5--7.6$\times$ over the strongest GPU baseline. It also accelerates
index construction by 2.4--14.8$\times$ over the GPU baseline.
\end{abstract}

\maketitle

\pagestyle{\vldbpagestyle}
\begingroup\small\noindent\raggedright\textbf{PVLDB Reference Format:}\\
\vldbauthors. \vldbtitle. PVLDB, \vldbvolume(\vldbissue): \vldbpages, \vldbyear.\\
\href{https://doi.org/\vldbdoi}{doi:\vldbdoi}
\endgroup
\begingroup
\renewcommand\thefootnote{}\footnote{\noindent
This work is licensed under the Creative Commons BY-NC-ND 4.0 International License. Visit \url{https://creativecommons.org/licenses/by-nc-nd/4.0/} to view a copy of this license. For any use beyond those covered by this license, obtain permission by emailing \href{mailto:info@vldb.org}{info@vldb.org}. Copyright is held by the owner/author(s). Publication rights licensed to the VLDB Endowment. \\
\raggedright Proceedings of the VLDB Endowment, Vol. \vldbvolume, No. \vldbissue\ %
ISSN 2150-8097. \\
\href{https://doi.org/\vldbdoi}{doi:\vldbdoi} \\
}\addtocounter{footnote}{-1}\endgroup

\ifdefempty{\vldbavailabilityurl}{}{
\vspace{.3cm}
\begingroup\small\noindent\raggedright\textbf{PVLDB Artifact Availability:}\\
The source code, data, and/or other artifacts have been made available at \url{\vldbavailabilityurl}.
\endgroup
}

\section{Introduction}
With the proliferation of recommendation systems and AI technologies, similarity search over embedding vectors has become a fundamental task. Embedding techniques map diverse data objects, such as documents, images, and audio, into high-dimensional vectors that effectively capture their complex semantic information~\cite{bert, ImageNet, wav2vec}. Since semantic similarity can be measured by proximity in the vector space, similarity search can be implemented through vector search and is commonly formulated as approximate nearest neighbor search (ANNS), which retrieves the $k$ approximate nearest neighbors ($k$-ANN) of a query vector from a collection of vectors derived from the original objects.
In practical scenarios, similarity search is often evaluated together with structured
predicates, such as retrieving visually similar products within a price range or videos within a time interval. This motivates range-filtering approximate nearest neighbor search (RFANNS), which returns $k$-ANN of a query vector among objects whose numerical attributes fall in a query range.

\begin{figure}[t]
  \centering
    \begin{minipage}{0.48\linewidth}
      \centering
      \includegraphics[width=\linewidth]{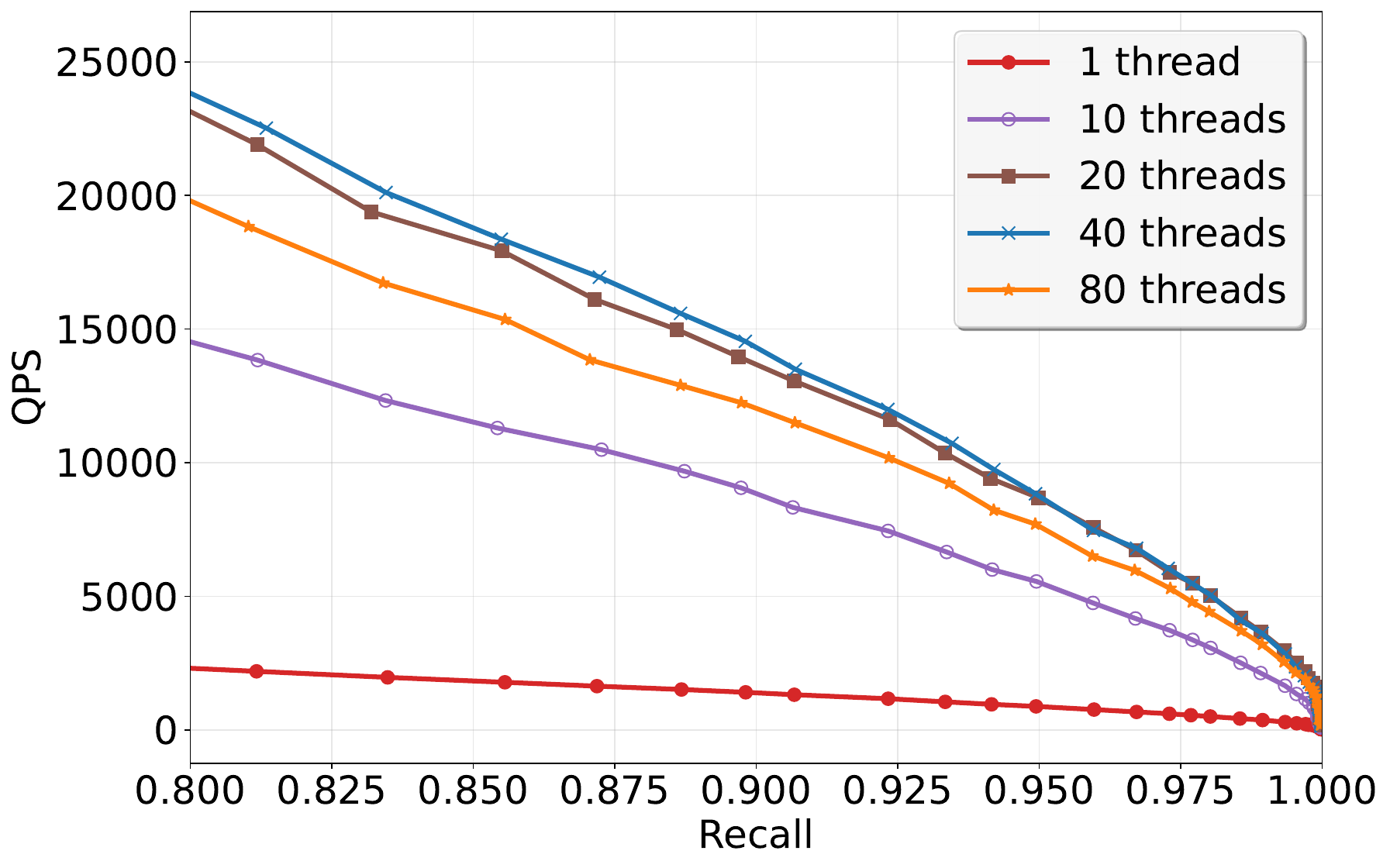}
      \captionof{figure}{Search performance of WoW~\cite{WoW} on GIST under different thread counts.}
      \label{fig:threads-wow}
    \end{minipage}
  \hfill
    \begin{minipage}{0.48\linewidth}
      \centering
      \includegraphics[width=\linewidth]{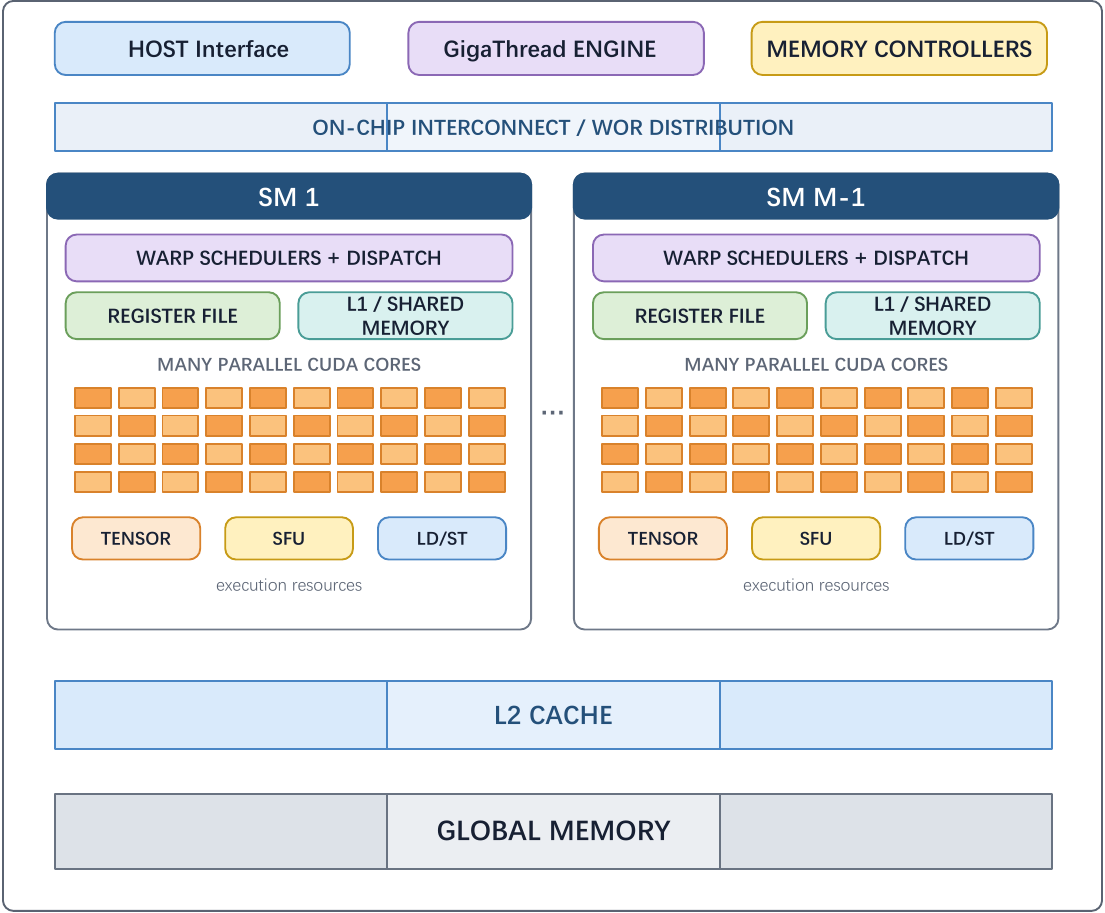}
      \captionof{figure}{GPU Architecture}
      \label{fig:Arch}
    \end{minipage}
\end{figure}

Existing RFANNS indexes~\cite{SeRF, DSG, UNIFY, WST, iRG, WoW, DIGRA} have mostly been designed as dedicated proximity graph index for CPU execution and have significantly improved search quality and index organization.
However, CPU-based approaches face system-level bottlenecks in modern high-throughput vector-database serving, even when parallelized across multiple cores. In particular, their performance is constrained by memory bandwidth, cache contention, limited register resources. As shown in \autoref{fig:threads-wow}, simply increasing the CPU thread count yields diminishing marginal gains on WoW~\cite{WoW}, a representative RFANNS index, as contention for shared resources intensifies. This observation suggests that further CPU-side optimization alone is unlikely to provide scalable throughput, motivating a GPU-based solution.

GPUs provide massive data-parallel computing capability and can substantially accelerate RFANNS, whose runtime is dominated by compute-intensive distance evaluations. Yet RFANNS is not a straightforward GPU porting problem: it requires both highly parallel algorithm design and hardware-friendly implementation. Directly mapping CPU-oriented methods to GPUs is often inefficient, as such methods typically fail to exploit GPU architectural characteristics.
Moreover, existing GPU approaches also fall short.
Generic filtering strategies built on unconstrained GPU ANNS are highly selectivity dependent: post-filtering wastes work when few retrieved candidates pass the range predicate, in-filtering can break valid-subset connectivity, and pre-filtering lacks an effective query-specific graph. Specialized GPU RFANNS indexes, such as Garfield~\cite{Garfield}, partition data into attribute cells and search across cell-local graphs, but this local design can repeatedly enter slow convergence phases and can compute distances to many invalid neighbors when a query range only partially overlaps a cell.

The key difficulty of processing RFANNS on GPUs is that the index to be used is determined at query time. For a given range, the ideal structure is the proximity graph built directly over the valid subset, which we call the Oracle PG. However, since the valid subset varies across queries, the Oracle PG cannot be precomputed for all possible ranges or reconstructed online in its entirety. An RFANNS index should therefore approximate the query-relevant portion of the Oracle PG, especially the expansion neighbors of vertices reached during greedy search. This observation suggests the key requirements for an effective GPU RFANNS method: providing high-quality range-aware expansion neighbors, enabling their online identification under arbitrary ranges, and organizing them in a memory layout with a query processing procedure that supports efficient GPU execution.

In this paper, we propose FROG, a GPU-oriented RFANNS index built around a globally aware, vertex-centric design. FROG stores range-diverse expansion-neighbor candidates for each vertex and identifies expansion neighbors across the search path online for given query ranges. Together with fusion-based candidate selection, bottom-up parallel construction, and SIMT-aware query processing, FROG makes the design efficient on GPUs.
Our contributions are summarized as follows.

\ding{202} Based on the analysis of the RFANNS problem and existing solutions, we propose a globally aware, vertex-centric design philosophy, as presented in \autoref{sec:motivation}.

\ding{203} We introduce a GPU-friendly index structure for FROG and design a fusion-distance-based candidate selection criteria, which are detailed in \autoref{sec:index_structure} and \autoref{sec:enc_selection}, respectively.

\ding{204} We develop GPU-efficient algorithms and implementations for FROG, including a bottom-up parallel construction algorithm (\autoref{sec:index_construction}) and a SIMT-aware query processing procedure (\autoref{sec:query_processing}) with hotspot localization, control-flow compaction, and lazy duplicate checking.

\ding{205} We conduct extensive experiments on six datasets in \autoref{sec:experiments}, showing that FROG achieves robust throughput across selectivities and accelerates both query processing and index construction over CPU and GPU baselines.

\section{Preliminaries}
\label{sec:preli}

\subsection{Problem Formulation}
\label{sec:preli_formu}
$k$-ANNS represents the canonical unconstrained formulation of vector search and is defined as follows.

\begin{definition}[$k$-ANNS]
\label{def:k-anns}
Given a vector set $V\subseteq\mathbb{R}^d$, a query vector $v_q\in\mathbb{R}^d$, and a positive integer $k$, $k$-ANNS approximately returns the top-$k$ vectors from $V$ ranked by their distances to $v_q$.
\end{definition}

In this paper, we study a constrained variant of $k$-ANNS, i.e., range-filtering $k$-ANNS (RFANNS).
Let $O=\{o_i=(v_i,a_i)\mid v_i\in\mathbb{R}^d,\ a_i\in\mathbb{R}\}_{i=1}^{n}$ be a set of objects, where $v_i$ and $a_i$ denote the vector and numerical attribute of object $o_i$, respectively. We use $V=\{v_i\mid o_i\in O\}$ and $A=\{a_i\mid o_i\in O\}$ to denote the corresponding vector set and attribute-value set, respectively. 
For two objects $o_i=(v_i,a_i)$ and $o_j=(v_j,a_j)$, we define their vector distance as the L2 distance $\delta_v(o_i,o_j)=\lVert v_i-v_j\rVert_2$. As detailed in \autoref{sec:index_structure}, attribute values are mapped to an ordered rank domain before indexing; throughout the subsequent analysis and implementation, $\delta_a(o_i,o_j)$ denotes the normalized L1 distance between their attribute ranks.
A query is represented as $q=(v_q,[l_q,r_q],k)$, where $v_q\in\mathbb{R}^d$ is the query vector, $[l_q,r_q]\subseteq\mathbb{R}$ is the query range, and $k$ is the requested number of results. The objects satisfying the range predicate are denoted by $O_{[l_q,r_q]}=\{o_i\in O\mid a_i\in[l_q,r_q]\}$, and their corresponding vector set is $V_{[l_q,r_q]}=\{v_i\mid o_i\in O_{[l_q,r_q]}\}$.
The selectivity of a range-filtering query is defined as $\sigma=|O_{[l_q,r_q]}|/|O|$, which measures the fraction of objects that pass the filter.

We present the formal definition of the RFANNS problem as follows. 
\begin{definition}[RFANNS]
\label{def:rfanns}
Given an object set $O$ and a query $q=(v_q,[l_q,r_q],k)$, RFANNS approximately returns the top-$k$ objects from $O_{[l_q,r_q]}$ ranked by the vector distance $\delta_v$ between their corresponding vectors and $v_q$.
\end{definition}


\subsection{Proximity Graph for $k$-ANNS}
\label{sec:preli_pg}
Recent studies~\cite{DPG,survey2021,AziziEP25} report that proximity-graph-based indexes have become the dominant paradigm for vector search, where PG vertices represent high-dimensional vectors and PG edges connect nearby neighbors in the vector space.
Representative PGs include relative-neighborhood-graph (RNG)-like graphs~\cite{NSG,NSSG,DPG}, navigable small-world graphs~\cite{NSW,HNSW}, and Vamana~\cite{DiskANN}.
Despite differences in edge-set construction strategies, these methods generally rely on a broadly similar greedy graph traversal paradigm, as illustrated in Algorithm~\autoref{alg:knn_search}.

\begin{algorithm}[t]
\small
\caption{\textsc{KANNSearch}$(G,v_q,k,\mathrm{ef_s},ep)$}
\label{alg:knn_search}
\begin{algorithmic}[1]
    \Require PG $G$, query $v_q$, result size $k$,
             search width $\mathrm{ef_s}$, and entry point $ep$
    \Ensure The $k$-ANN of $v_q$
    \State $i \gets 0$
    \State $pool[0] \gets (ep,\operatorname{dist}(v_q,ep))$
    \While{$i < \mathrm{ef_s}$}
        \State $u \gets pool[i]$
        \ForAll{$v \in N_G(u)$}
            \State Insert $(v,\operatorname{dist}(v_q,v))$ into $pool$
        \EndFor
        \State Sort $pool$ and retain the $\mathrm{ef_s}$ closest vertices
        \State $i \gets$ index of the first unexpanded vertex in $pool$
    \EndWhile
    \State \Return $pool[0,\ldots,k-1]$
\end{algorithmic}
\end{algorithm}

Let $G=(V, E)$ denote a PG, where $V$ indicates the vertex set and $E$ the edge set. For each vertex $u \in V$, we use $N_G(u)$ to denote its neighbor set in $G$. The search starts from an entry point $ep$, which is inserted into a candidate pool $pool$ maintained in ascending order of distance to the query $v_q$ (Lines 1–2). The pool retains at most the current $\mathrm{ef_s}$ closest candidates. The algorithm then repeatedly selects the closest unexpanded vertex $u$ from $pool$ and expands it until no unexpanded vertex remains among the retained $\mathrm{ef_s}$ candidates (Lines 3–8). During the expansion of $u$, each neighbor $v \in N_G(u)$ is considered as a potential $k$-ANN candidate (Line 5). The algorithm computes $\operatorname{dist}(v_q,v)$ and inserts the resulting vertex–distance pair into $pool$ (Line 6). After all neighbors of $u$ have been evaluated, $pool$ is reordered by distance and truncated to retain only the $\mathrm{ef_s}$ closest vertices (Line 7). The index $i$ is then updated to the position of the closest unexpanded vertex, which will be expanded in the next iteration (Line 8). Once all retained candidates have been expanded, the algorithm returns the first $k$ vertices in $pool$ as the approximate $k$ nearest neighbors of $v_q$ (Line 9).

In addition to ANNS methods that directly provide static graphs, mainstream RFANNS methods~\cite{SeRF, DSG, UNIFY, WST, iRG, WoW, DIGRA} also use PGs and greedy search as the fundamental building blocks.

\subsection{GPU Characteristics}
\label{sec:preli_gpu}
GPUs are designed to provide high throughput by executing a large number of lightweight threads in parallel. GPU threads are organized hierarchically into grids, thread blocks, and warps. A thread block is scheduled onto a streaming multiprocessor (SM), while a warp, which typically consists of a fixed number of threads, is the basic execution unit within an SM. Under the single-instruction, multiple-thread (SIMT) execution model, threads in the same warp execute the same instruction in a lockstep manner. GPUs also employ a hierarchical memory system comprising per-thread registers, per-block shared memory, and high-latency global memory.

These architectural characteristics impose several performance considerations. GPU utilization requires sufficient independent tasks and multiple active warps per SM to hide memory and dependency stalls; sequential dependencies and frequent synchronization therefore limit performance. Moreover, performance benefits from reducing global-memory traffic and coalescing accesses by threads in the same warp. When threads in a warp follow different control-flow paths, warp divergence forces those paths to execute serially and reduces effective parallelism. GPU algorithms should thus minimize data-dependent branching or compact active threads.

\section{Motivation}
\label{sec:motivation}
\label{problem_analysis}
For the unconstrained ANNS, a PG is constructed over the dataset during the index construction phase. In the PG, each object is associated with a set of neighbors, namely the vertices connected to it by edges.
In the query-processing phase, Algorithm~\autoref{alg:knn_search} performs a sequence of hops over the PG toward vertices closer to $q$; at each hop, the selected vertex is expanded by evaluating all of its neighbors.
We refer to the neighbors whose distances to the query vector are actually computed during an expansion as \textbf{expansion neighbors (ENs)}.

By contrast, the ENs of an expanded vertex in RFANNS are not equivalent to the neighbors selected from the entire dataset for unconstrained ANNS.
RFANNS produces results from an object subset $O_{[l_q,r_q]}$ that cannot be determined in advance, because the attribute constraint is specified only at query time.

\begin{figure}[t]
  \centering
  \includegraphics[width=\linewidth]{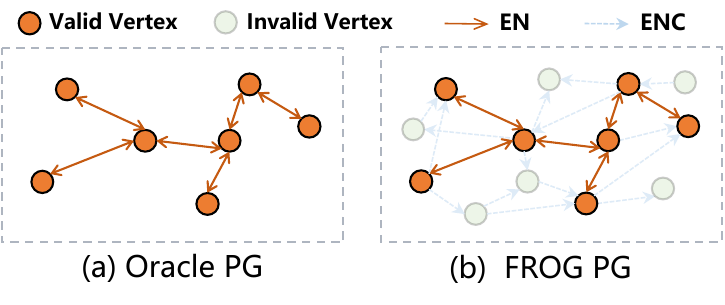}
  \caption{Oracle PG and FROG PG built on valid subset.}
  \label{fig:idealPG}
\end{figure}

Ideally, an RFANNS query would be performed on the \textbf{Oracle PG} shown in \autoref{fig:idealPG}(a), defined as the native PG constructed over the vector subset $V_{[l_q,r_q]}$ corresponding to the valid object subset $O_{[l_q,r_q]}$, thereby reducing the problem to an unconstrained ANNS task. In other words, RFANNS can achieve near-ideal query performance if it can efficiently identify a high-quality PG over the valid subset. Since not all vertices in the PG are expanded during graph traversal, an equivalent minimal query-relevant structure includes the ENs of the vertices traversed along the search path.
Therefore, an RFANNS method should pursue the following design goals: \textbf{1) ensuring near-ideal EN quality} by making the identified ENs closely approximate those of the Oracle PG, as shown in \autoref{fig:idealPG}(b); \textbf{2) rapidly identifying ENs online} under arbitrary query-time range constraints; and, if designed for GPUs, \textbf{3) enabling GPU-friendly execution} through an index structure amenable to parallel computation.

\subsection{Limitations of Existing Works}
\label{sec:existing_limitations}
Existing solutions can be broadly classified into three main categories, none of which adequately balances the above objectives. To better understand these limitations, we next examine each category in detail.

\subsubsection{Generic GPU-Based Methods}
Generic filtering strategies apply unconstrained GPU ANNS machinery to RFANNS. Post-filtering searches a PG over the entire database and filters an enlarged candidate set afterward; at low selectivity, obtaining enough valid results can require an impractically large candidate set. In-filtering skips invalid vertices on the same PG, disrupting query-valid connectivity and degrading recall. Pre-filtering first materializes the valid subset but cannot construct an effective query-specific index online, leaving brute-force distance computation as the practical fallback. Consequently, these strategies are competitive only in limited selectivity or workload regimes and do not consistently satisfy Objectives 1 and 2.

\subsubsection{Specialized GPU-Based Methods} 
As the only known GPU-tailored RFANNS method, Garfield~\cite{Garfield} partitions the database into multiple cells according to specific attributes, independently builds a PG for each cell, and constructs cross-cell edges. Nevertheless, its design remains suboptimal in several aspects.
On the one hand, Garfield incurs additional overhead due to its long search trajectories. By sequentially traversing related cells and performing greedy search within each, its trajectory becomes a concatenation of search paths over multiple cell-induced subgraphs. Although optimizing the entry point of each subsequent cell saves a few hops, it cannot avoid the slowdown of greedy search near convergence. Garfield thus repeatedly undergoes this inefficient convergence phase, leading to a longer search path and lower query throughput.
On the other hand, when the intersection between the query attribute range and the interval corresponding to a cell is small, many neighbors evaluated during the search are invalid, causing substantial computational resources to be wasted on unnecessary distance computations. 

\subsubsection{Straightforward GPU Ports of Typical CPU-Based Methods}
Typical CPU-oriented designs also map poorly to GPUs. SeRF and DSG require each expanded vertex to filter many tagged edges, amplifying global-memory traffic. iRG composes query-specific graphs from segment-local PGs, which may deviate from the Oracle PG, and its query processing procedure exposes insufficient GPU parallelism. WoW relies on sequential insertion during construction and is therefore incompatible with efficient GPU execution. Other CPU methods exhibit similar limitations in memory access, control flow, or parallelism.

\subsection{From Local Optimum to Global Awareness}
FROG is designed around a global-awareness philosophy. Many existing indexes are constructed as combinations of multiple locally optimal substructures. For example, iRG and Garfield, representative methods on CPU and GPU platforms, respectively, build a number of subgraphs that achieve optimal query performance within individual segments or cells. However, combining locally optimal substructures—whether by reselecting edges from them or by connecting multiple subgraphs through search paths—does not guarantee the construction of a globally optimal PG.

\textbf{In contrast, FROG does not aim for optimality within any particular local substructure. Instead, it is designed to support the generation of the native PG for the valid subset, reflecting a globally aware design philosophy.}

Accordingly, the content stored in the FROG index is vertex-centric rather than the prevailing subgraph-centric form. Specifically, FROG pre-stores, for each vertex, a candidate set that contains a sufficient number of high-quality potential ENs for all possible query attribute ranges; the elements in this set are referred to as \textbf{expansion neighbor candidates (ENCs)}. Then, during the query-processing phase, FROG selects ENs from this set online, aiming to reconstruct the ENs of the Oracle PG as closely as possible.

\section{Index Structure}
\label{sec:index_structure}
As mentioned earlier, the contents stored in the FROG index are the ENCs of individual dataset objects. This section further explains how FROG organizes ENCs.

\begin{figure}[t]
    \centering

    \begin{subfigure}[t]{0.5\linewidth}
        \centering
        \includegraphics[width=\linewidth]{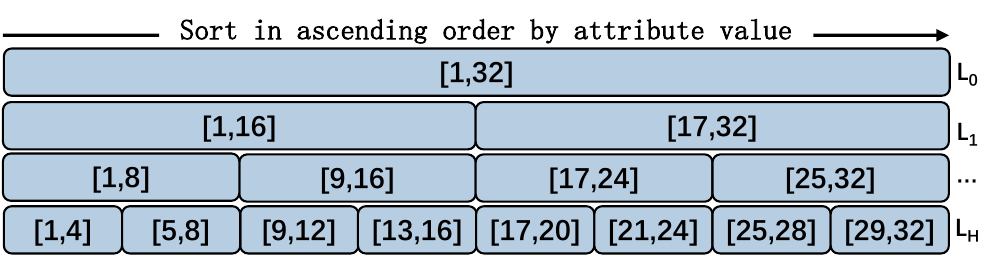}
        \caption{Logic Structure}
        \label{fig:logic_structure}
    \end{subfigure}
    \hfill
    \hspace{-2mm}
    \begin{subfigure}[t]{0.5\linewidth}
        \centering
        \includegraphics[width=\linewidth]{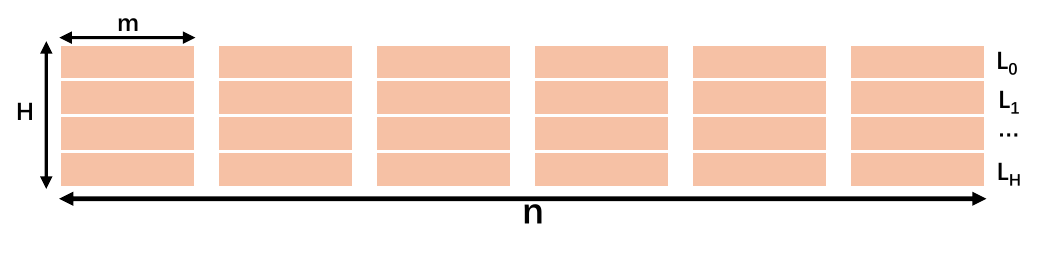}
        \caption{Physical Structure}
        \label{fig:physical_structure}
    \end{subfigure}
    
    \caption{Index Structure of FROG.}
    \label{structure}
\end{figure}

\autoref{structure} presents both the logical organization and the physical layout of FROG. FROG first sorts objects in ascending order of attribute value and assigns them ranks in $[0,n-1]$. It maintains the correspondence between ranks and original attribute values; query endpoints are converted by lower- and upper-bound lookup, so an attribute range is represented by the contiguous rank interval containing exactly its valid objects. The index is then organized in the resulting \emph{attribute-rank domain}.

As illustrated in \autoref{fig:logic_structure}, FROG builds a segment-tree-style hierarchy on this domain. Segment trees are classical interval structures introduced by Bentley~\cite{bentley1977solutions} and later adapted to RFANNS by $\beta$-WST~\cite{WST} and iRG~\cite{iRG}; FROG redesigns their contents and layout for GPU-oriented, vertex-centric EN identification. The root covers the full rank domain, and each internal node represents a contiguous rank interval that is recursively split into two equal subintervals. Nodes in the same layer therefore partition the domain, while deeper layers provide progressively finer range granularities.

Rather than storing a locally optimized PG for a segment, FROG associates each segment-tree node with a set of vertex-centric ENCs: for every object covered by the node, it stores $m$ ENCs selected from that node's segment. This relationship is materialized by the dense array in \autoref{fig:physical_structure}. In particular, the cell in column $i$ and row $j$ represents the $m$ ENCs of object $o_i$ at layer $L_j$; these ENCs are selected from the segment containing $o_i$ at layer $L_j$ in \autoref{fig:logic_structure}. Thus, each object has an independent fixed-size block of $H\times m$ entries, and the blocks of all $n$ objects form an $n\times H\times m$ memory region. Introducing this layout together with the logical tree makes explicit that the tree organizes the candidate scope, whereas the physical index remains vertex-centric.

FROG chooses the segment-tree hierarchy for two reasons. First, its layers cover query ranges at different granularities, from extremely selective to highly unselective predicates. During the query-processing phase, the hotspot localization rule in \autoref{sec:hotspot} activates layers whose interval scales match the query range, allowing online EN identification to adapt to selectivity. Second, intervals within a layer are disjoint, so each object belongs to exactly one node at that layer. Construction tasks for different nodes consequently share neither objects nor output regions, eliminating inter-node synchronization conflicts and enabling parallel GPU construction.

Although segment-tree-based RFANNS indexes have proved effective on CPUs, FROG differs from them in two important respects. The first is the vertex-centric ENC content described above. The second is that the number of retained layers is tunable. The root still covers all objects, but recursion may stop before singleton leaves, so a leaf can contain multiple consecutive objects. Omitting the deepest, rarely activated layers improves space efficiency under practical selectivities. We use $n_{\mathrm{inv}}$ for the number of layers omitted from a fully expanded tree and $H$ for the depth of the retained tree.

The contiguous physical layout supports vectorized loads and coalesced GPU memory accesses. FROG also avoids explicitly storing segment-tree metadata: when construction or search requires a segment boundary or child interval, it reconstructs that information from the parent using a deterministic splitting rule. The segment tree therefore acts as a lightweight logical scaffold, while the stored index remains compact and GPU-friendly.

\section{ENC Selection Criteria}
\label{sec:enc_selection}
Although FROG selects ENCs within individual segments, it evaluates them by the quality of the ENs they can produce at query time. \autoref{sec:suboptimal_pg} analyzes an idealized vector-only segment method to expose the inherent boundary effect of segment-local optimization; \autoref{sec:fusion_criterion} then introduces FROG's fusion-distance criterion.

\subsection{Suboptimality of Locally Optimized PG}
\label{sec:suboptimal_pg}
To isolate the structural limitation of segment-local optimization, consider an idealized naive method that, at every segment-tree node, connects each object to its top-$m$ vector-nearest objects within that segment. This method is deliberately stronger than a practical approximate construction, but its segment-local optimum need not yield high-quality ENs for a query range.

For a closed-form analysis, assume that vector distances are continuous and independent of attribute ranks. Consider a segment $I=[L,R]$ of length $\ell=R-L+1$. For $o_i\in I$, let $t=i-L$ be its offset. Under the independence assumption, the attribute ranks of its top-$m$ vector neighbors are uniformly distributed over $I\setminus\{i\}$. Hence, the expected attribute-rank distance of a selected candidate is
\begin{equation}
D_{\ell}(t)
=
\frac{
t(t+1)+(\ell-1-t)(\ell-t)
}{
2(\ell-1)
} .
\label{eq:naive_attr_dist}
\end{equation}
For a boundary object, $D_{\ell}^{\mathrm{bd}}=\ell/2$. For a center object in an odd-length segment, $D_{\ell}^{\mathrm{ct}}=(\ell+1)/4$, so $D_{\ell}^{\mathrm{bd}}/D_{\ell}^{\mathrm{ct}}=2\ell/(\ell+1)\rightarrow 2$. Thus, a boundary object has almost twice the expected candidate attribute distance of a center object. The two central positions of an even-length segment give the same asymptotic result.

Averaging uniformly over object positions gives $\mathbb{E}_{t}[D_{\ell}(t)]=(\ell+1)/3$. As a boundary-free reference, consider an ideal structure that supplies every object with a same-length attribute window centered at that object, rather than a window fixed by segment boundaries. Such an object-adaptive structure is difficult to materialize for arbitrary query-time ranges, but it represents the candidate locality available without artificial segment boundaries. Its expected attribute distance is $(\ell+1)/4$ for odd $\ell$. The ratio is therefore $4/3$, i.e., fixed segment boundaries introduce a $33.3\%$ expected attribute-distance inflation under this model.

The independence assumption makes these expectations explicit, whereas uniform raw attributes are unnecessary. Mapping both attribute values and query endpoints to the ordered rank domain preserves range membership, so the boundary analysis applies to arbitrary empirical marginal attribute distributions. Correlation between attributes and vectors may, however, change the exact expectations.

This inflation further reduces the number of ENCs that survive range filtering. To see this, sample a query interval uniformly from all non-empty intervals containing rank $i$. A direct interval-counting argument shows that the probability of another rank $j$ appearing in the same interval decreases as $j$ moves away from $i$ on either side. Thus, the larger candidate attribute distance caused by fixed segment boundaries yields fewer valid ENCs in expectation.

Finally, this shortage worsens the expected vector distance of ENs. Let $\ell_\lambda$ be the segment length at level $\lambda$, let $s_\lambda=\ell_\lambda-1$, and use the CDF value of vector distance only as an analytical rank normalization. The expected normalized distance of the $r$-th nearest object among $s_\lambda$ objects is $r/(s_\lambda+1)$. Hence, the average normalized vector distance of the top-$m$ candidates at level $\lambda$ is
\begin{equation}
\mu_\lambda
=
\frac{1}{m}
\sum_{r=1}^{m}
\frac{r}{s_\lambda+1}
=
\frac{m+1}{2(s_\lambda+1)} .
\label{eq:level_vector_dist}
\end{equation}
Since deeper levels have smaller segments, $\mu_\lambda$ increases with depth. If higher-level candidates do not provide enough valid ENs, the search must draw more ENs from deeper levels and therefore incurs a larger expected vector distance.

\subsection{ENC Selection Criterion Based on Fusion Distance}
\label{sec:fusion_criterion}

Unlike prior RFANNS methods that rely solely on vector distance, FROG also takes the normalized L1 attribute distance $\delta_a$ defined in \autoref{sec:preli_formu} into account. For example, when selecting ENCs for an object \(o\) located near a segment boundary, a neighbor \(x\) in an adjacent segment that is closer in the attribute space but slightly farther in the vector space is a better choice than another neighbor \(y\) located near the opposite boundary of the same segment. This is because objects with smaller attribute distances are more likely to fall into the same query range as \(o\). When query ranges are small, such objects may become ENs even without a global advantage in vector distance. Conversely, objects that are far from \(o\) in the attribute space can stand out only if they are exceptionally close in the vector space, since larger query ranges intensify the competition under a fixed EN budget.

FROG is based on the observation that, when selecting ENCs for \(o=(v,a)\), objects with smaller attribute distance to \(o\) can tolerate larger vector distances. Therefore, \textbf{FROG adopts a fusion distance \(\delta_f\), which jointly incorporates attribute distance \(\delta_a\) and vector distance \(\delta_v\), as the selection criterion for ENCs.} Specifically, the fusion distance between two objects \(x\) and \(y\) is defined as
\begin{equation}
    \delta_f(x,y)
    =
    \delta_v(x,y)
    \left[
        1-\beta\left(1-\delta_a^{\gamma}(x,y)\right)
    \right],
    \label{eq:fusion-distance}
\end{equation}
where \(\beta \in (0,1)\) and \(\gamma > 0\) are hyperparameters. The parameter \(\beta\) determines the maximum shrinkage applied to the vector distance, while \(\gamma\) controls how rapidly the effect of attribute distance decays.

As the attribute distance decreases, the fusion distance decreases accordingly. As a result, objects that would be blocked by a segment boundary under the naive method can become competitive and be selected indirectly at a higher layer of FROG.

Let $X=\delta_v$ denote the actual vector distance and let $Z=\delta_a$ denote normalized attribute-rank distance. Write the fusion distance as $Y=XW(Z)$, where $W(Z)=1-\beta+\beta Z^\gamma$. Since $W(Z)$ increases with $Z$, attribute-closer objects receive smaller fusion distances and become more likely to be selected.

Consider a candidate pool of size $s$ from which $m$ ENCs are retained, where $q=m/s\ll 1$. Assume that $X$ and $Z$ are independent and that the lower tail of the vector-distance CDF satisfies $F_X(x)=c_Xx^\alpha+o(x^\alpha)$ for some $c_X,\alpha>0$. Since $\Pr[Y\le\tau\mid Z=z]=F_X(\tau/W(z))$, rare selection reweights attribute distances in proportion to $W(Z)^{-\alpha}$. Let $C_\gamma=\operatorname{Cov}(Z,Z^\gamma)>0$. For small $\beta$,
\begin{equation}
\mathbb{E}_{\beta}[Z]
=
\mathbb{E}[Z]
-
\alpha\beta C_\gamma
+
O(\beta^2).
\label{eq:fusion_attr_first_order}
\end{equation}
For example, $C_\gamma=\gamma/[2(\gamma+1)(\gamma+2)]$ when $Z\sim U(0,1)$. Therefore, fusion distance reduces the expected attribute distance of ENCs by a first-order term in $\beta$.

Because range-survival probability decreases with $Z$, this first-order reduction increases valid-ENC availability. For example, under the local approximation $p(Z)\approx1-cZ$, the increase is $c\alpha\beta C_\gamma+O(\beta^2)$.

The direct vector-distance cost is only second order. Let $V_\gamma=\operatorname{Var}(Z^\gamma)$. Under the same rare-selection regime,
\begin{equation}
\frac{\mathbb{E}_\beta[X\mid\mathrm{sel}]}
     {\mathbb{E}_0[X\mid\mathrm{sel}]}
=1+\frac{\alpha+1}{2}\beta^2V_\gamma+O(\beta^3).
\label{eq:fusion_vector_cost}
\end{equation}
Hence, fusion selection trades a first-order gain in valid-ENC availability for only a second-order inflation in the vector distance of selected candidates.

Fusion selection is therefore expected to obtain more valid ENCs from higher levels and reduce the need for deeper-level ENs with larger $\mu_\lambda$. This first-order/second-order tradeoff predicts a beneficial small-$\beta$ regime, but does not by itself prove end-to-end dominance over a practical graph method.

Equivalently, for two candidates $x$ and $y$, FROG ranks $x$ before $y$ if $X_x/X_y < (1-\beta+\beta Z_y^\gamma)/(1-\beta+\beta Z_x^\gamma)$. When $Z_x<Z_y$, the right-hand side exceeds $1$. Thus, an attribute-closer candidate can tolerate a moderately larger vector distance, which is exactly the desired behavior for RFANNS.

\section{Index Construction}
\label{sec:index_construction}

\subsection{Construction Procedure}

\begin{figure}[b]
  \centering
  \includegraphics[width=\linewidth]{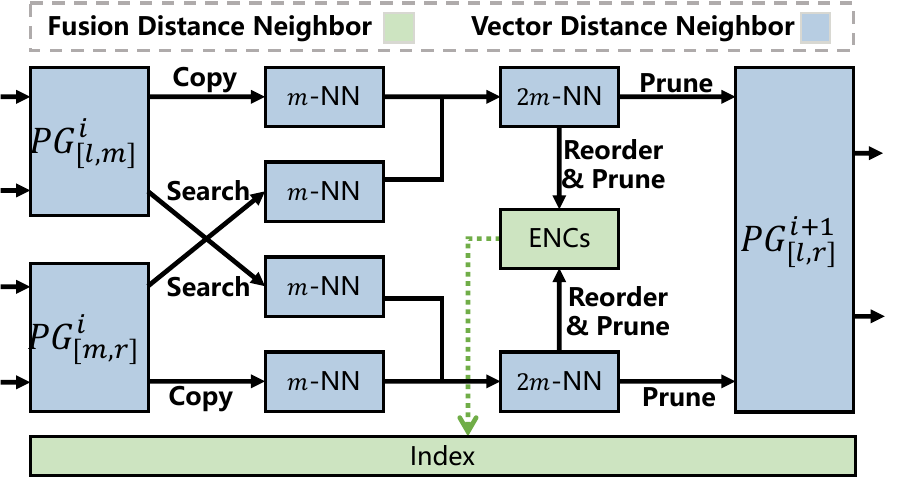}
  \caption{The inductive step of FROG construction.}
  \label{fig:construction}
\end{figure}

FROG adopts a bottom-up, intra-layer concurrent construction paradigm.
Within each segment of FROG, the ENCs depend on the intermediate results of its two child segments, thereby reusing previous computations and reducing overhead. Starting from the leaf nodes, the construction proceeds upward layer by layer until reaching the root.
Since each layer is uniformly partitioned into disjoint segments, and the corresponding objects are mutually exclusive, the subtasks within the same layer can be executed independently, providing a natural basis for parallelism.

The base case corresponds to leaf nodes, where ENCs are computed using an efficient brute-force kernel. Within each leaf node, every vertex sequentially computes its fusion distance to the other vertices and updates its $m$ nearest neighbors accordingly.

The subsequent nodes involve data at a much larger scale, making it prohibitively expensive to compute the fusion distance between every pair of objects. If previously computed results can be effectively reused, the overall computational cost can be significantly reduced.

However, unlike iRG, which relies solely on vector distances and can directly use the PGs of child segments as ANNS indexes for neighbor retrieval in the parent segment, the ENCs stored in FROG segments do not serve as a natural index. This is because fusion distance is not a metric: in general, it does not satisfy identity of indiscernibles or the triangle inequality, thereby undermining the foundation of PG navigability. Traversing a graph built on fusion distance leads to poor performance.

To enable similarity to be propagated across layers, FROG employs vector distance as an intermediate proxy. It first retrieves a neighbor set of size $m'$ ($m' > m$) under vector distance, computes the fusion distances to these neighbors, and then re-ranks them according to fusion distance. The $m$ nearest elements after re-ranking are selected as ENCs. This design is based on the intuition that points that are far apart in vector space are less likely to be close under fusion distance.

To accelerate the retrieval of vector-distance neighbors, FROG still constructs a temporary PG based on vector distances, although this PG is not persisted in the final index. The temporary PGs of a child layer are built while that layer is constructed and remain available only until their parent layer has been completed. Their storage is then reused in place: the newly constructed parent PG overwrites the child PGs rather than requiring a separate allocation or an explicit deallocation. We denote the candidate-pool width used by each construction-time greedy search on this temporary PG by $\mathrm{ef_c}$. With a carefully engineered design, the overhead introduced by this additional temporary index is virtually negligible.

Above, the inductive construction step, which generates the current segment from child segments at one layer lower, is shown in \autoref{fig:construction}.
For each object in $L_i$, the temporary PG of the child segment containing that object directly provides $m$ approximate nearest neighbors ($m$-ANN). Meanwhile, using the object itself as the query, Algorithm~\autoref{alg:knn_search} is performed on the temporary PG of the sibling segment in $L_{i+1}$.
The results are then merged to obtain an ANN set consisting of $2m$ neighbors, which serves two purposes. First, RNG pruning is applied to this set to construct the PG of the current segment, which is used as the index for the subsequent $L_{i-1}$ layer. Second, the fusion distances between the elements in this set and the target object are computed; the elements are then re-ranked and pruned based on these fusion distances, yielding the ENCs of the current layer, which are stored in the index.

\subsection{Implementation}
\subsubsection{Parallelism}
FROG exploits parallelism at both the segment and vertex levels. Since segments within the same layer are mutually independent, they can be constructed concurrently, and synchronization is only required between adjacent layers. 
Inside each segment, FROG maps one query vertex to one CUDA block.
For leaf nodes, the top-$m$ neighbors are maintained directly in registers, and insertion is implemented using \textit{ballot} and \textit{shfl}, avoiding shared-memory based priority queues.
For internal nodes, the candidate pool and expansion flags are kept in shared memory, while warp-level primitives are used to locate the next unexpanded candidate and update the pool efficiently.

\subsubsection{Early Termination}
Greedy search on the temporary PG consumes most of the construction runtime.
To accelerate it, FROG employs an early-termination technique inspired by~\cite{teofili2025patience}, since exact $m$-NN results are unnecessary.
First, search difficulty varies substantially: for most queries, the full neighbor set can be assembled quickly, while only a small fraction require many more iterations and some neighbors may never be reached due to graph-quality issues. Second, the closest neighbors who play a decisive role are typically discovered early. Early termination therefore tends to sacrifice only the tail neighbors, making the results still good enough.

Specifically, during search, we introduce a \textit{patience} parameter that specifies the number of consecutive non-improving iterations allowed. If none of the newly computed candidates enter the top-$m$ results, the patience counter is decremented by one; otherwise, it is reset to its initial value. The search terminates once the patience counter reaches zero.

\subsection{Cost Analysis}
\label{sec:build_cost}

We analyze the FROG-specific construction overhead relative to the
bottom-up segment-tree construction of iRG~\cite{iRG}. Let
$s_{\mathrm{leaf}}$ denote the maximum number of objects in a retained leaf
and let $H$ denote the number of retained layers.

\paragraph{Space Cost Analysis.}
The final FROG index stores $m$ ENCs for every object at every retained
layer and therefore occupies $O(nHm)$ space. During construction, FROG
additionally maintains temporary vector-distance PGs. At any layer, the
segments form a disjoint partition of the $n$ objects, so all temporary PGs
at that layer contain $O(nm)$ edges in total. Once a parent layer has been
constructed, its PGs overwrite the storage previously occupied by the two
child PGs. Consequently, construction never needs to retain temporary PGs
from multiple complete layers, and the additional construction workspace
for these PGs is $O(nm)$.

\paragraph{Time Cost Analysis.}
The leaf layer is constructed by brute-force comparison within each leaf.
Across all leaves, this requires $O(ns_{\mathrm{leaf}})$ pairwise fusion-
distance evaluations. At an internal layer, every object obtains $m$
neighbors directly from the PG of its own child segment and retrieves
another $m$ neighbors by searching the sibling PG. These $2m$ candidates
are also reused to construct the temporary parent PG, so FROG does not
introduce another graph search solely for ENC selection.

Relative to vector-distance construction, FROG only adds fusion-distance evaluation and re-ranking over $O(m)$ candidates per object. Since fusion-distance evaluation costs $O(m)$ and bitonic re-ranking costs $O(m\log^2 m)$, the additional work is $O(nm\log^2 m)$ per internal layer and $O(Hnm\log^2 m)$ over all retained layers. Including the leaf construction, the total FROG-specific construction work is $O(ns_{\mathrm{leaf}}+Hnm\log^2 m)$.

\section{Query Processing}
\label{sec:query_processing}

\subsection{Query Processing Procedure}
FROG approximates the Oracle PG online by reconstructing only its query-relevant portion. At each greedy-search hop, it identifies ENs for the current vertex from that vertex's pre-stored ENCs; the ENs accumulated along the search path form the required minimal structure.

\begin{algorithm}[t]
\small
\caption{\textsc{FROGSearch}$(ENC,v_q,l_q,r_q,k,\mathrm{ef_s},epn,H,N,B)$}
\label{alg:frog_search}
\begin{algorithmic}[1]
    \Require ENC of all vertices $ENC$, query vector $v_q$,
             query range $[l_q,r_q]$, result size $k$,
             search width $\mathrm{ef_s}$, number of entry points $epn$,
             segment tree depth $H$, total number of vertices $N$,
             expansion budget $B$
    \Ensure The $k$-ANN of $v_q$
    \State $(L_{start}, L_{end}) \gets \textsc{HotspotLocalization}(l_q,r_q,H,N)$
    \For{$j$ from $0$ to $epn-1$}
        \State Uniformly sample $ep_j$ from $[l_q,r_q]$
        \State $pool[j] \gets (ep_j,\operatorname{dist}(v_q,ep_j),0)$ \Comment{(id, dist, flag) of a vertex}
    \EndFor
    \State Sort $pool$ in ascending order of distance
    \State $i \gets 0$
    \While{$i < \mathrm{ef_s}$}
        \State $u \gets pool[i].id$
        \State $pool[i].flag \gets 1$
        \State $EN(u) \gets \textsc{ENIdentification}(ENC(u),u,l_q,r_q,L_{start},L_{end},B)$
        \State $new \gets \emptyset$ \Comment{clear for this iteration}
        \ForAll{$v \in EN(u)$}
            \If{\textsc{hashCheck}$(v)$} \Comment{lazy duplicate-evaluation check; true iff unevaluated}
                \State $new[v] \gets (v,\operatorname{dist}(v_q,v),0)$
            \EndIf
        \EndFor
        \State Sort $new$ in ascending order of distance
        \State $pool \gets \textsc{MergeTopEfS}(pool,new,\mathrm{ef_s})$ \Comment{maintain $\mathrm{ef_s}$ nearest neighbors}
        \State $i \gets$ the position of the first vertex in $pool$ with $flag=0$
    \EndWhile
    \State \Return $pool[0,\ldots,k-1]$
\end{algorithmic}
\end{algorithm}

Since queries are mutually independent, FROG assigns each query to a CUDA block, which is then scheduled by the CUDA runtime onto an SM for execution. Algorithm~\autoref{alg:frog_search} describes the execution of a single query $(v_q,[l_q,r_q],k)$.
The query processing procedure first invokes \textsc{HotspotLocalization} to obtain the actual ENC layers that will be used at query time, thereby reducing subsequent computational overhead (Line~1); further details are provided in \autoref{sec:hotspot}. It then uniformly samples $epn$ entry points from the query range $[l_q,r_q]$, computes their distances to $v_q$, and initializes the candidate pool with triples containing the vertex ID, distance, and expansion flag (Lines~2--4). After sorting the pool in ascending order of distance, the algorithm initializes the expansion pointer $i$ to the closest candidate (Lines~5--6).

\begin{figure}[t]
  \centering
  \includegraphics[width=\linewidth]{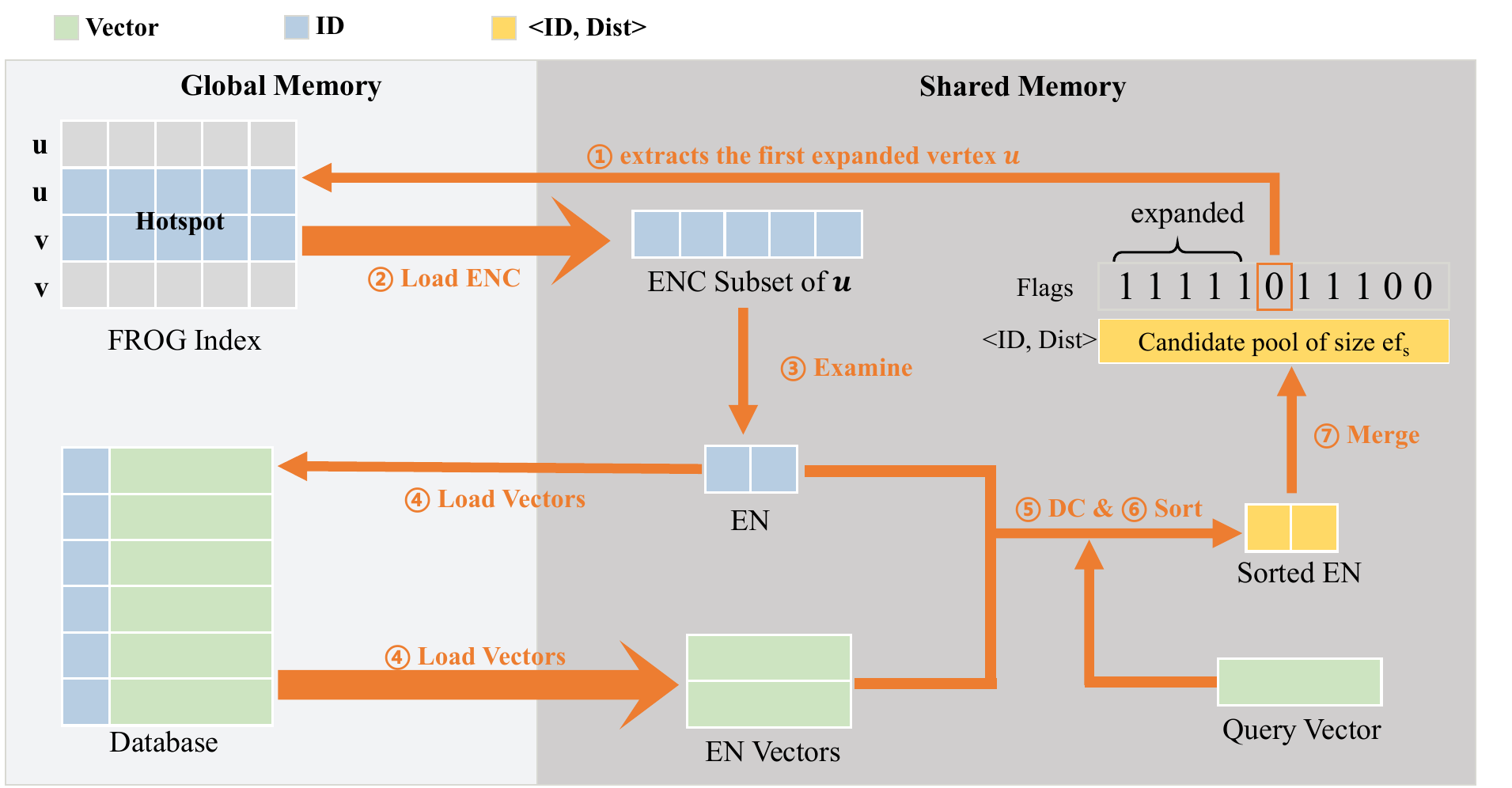}
  \caption{Workflow and memory layout of processing a query.}
  \label{fig:frog_search}
\end{figure}

The main loop follows the same best-first greedy expansion strategy as Algorithm~\autoref{alg:knn_search}, while adapting it to FROG's online EN identification. In each iteration, the algorithm extracts the first unexpanded vertex $u$ from $pool$ and marks its flag (Lines~7--9). It then calls \textsc{ENIdentification} to obtain the expansion neighbors $EN(u)$ of $u$ from the ENCs within the localized hotspot (Line~10) and clears the new-candidate array for the current iteration (Line~11). Before computing the distance of each EN, the algorithm performs \textsc{hashCheck} as a lazy duplicate-evaluation check at this point: it returns true only if the vertex has not previously been evaluated. Each such vertex is then inserted into the new-candidate array with an unexpanded flag (Lines~12--14). The new-candidate array is sorted by increasing distance (Line~15), merged with the current $pool$, and truncated to the top-$\mathrm{ef_s}$ closest vertices, thereby keeping the search frontier both ordered and bounded in size (Line~16). The pointer $i$ is updated to the first vertex in the pool whose flag remains zero; if no such vertex exists, $i$ becomes $\mathrm{ef_s}$ and the loop terminates (Line~17). Finally, the first $k$ entries of the pool are returned as the approximate nearest neighbors of $v_q$ (Line~18).

The data structures and data-loading process are illustrated in \autoref{fig:frog_search}. Global memory is shared across queries and stores the FROG index and original objects, while shared memory is private to each query (CUDA block) and holds the candidate pool, new-candidate array, query vector, etc. Each query only loads ENC entries within its hotspot, and only the vectors of ENs are fetched into shared memory for distance computation.

For the generic GPU primitives used in graph-based ANNS search, our implementation follows prior work~\cite{CAGRA, GANNS}. In particular, we adopt and adapt their designs for SIMT-based distance computation, parallel sorting, and erasable hash-table management. The detailed implementation and optimizations of FROG-specific operations, including \textsc{HotspotLocalization} and \textsc{ENIdentification}, are presented in the following subsections.

\subsection{Hotspot Localization}
\label{sec:hotspot}

\begin{algorithm}[t]
\small
\caption{\textsc{HotspotLocalization}$(l_q,r_q,H,N)$}
\label{alg:hotspot_localization}
\begin{algorithmic}[1]
  \Require query range $[l_q,r_q]$, segment tree depth $H$, total number of vertices $N$
  \Ensure localized ENC layer range $(L_{start},L_{end})$

  \State $L \gets 0,\ R \gets N-1,\ d \gets 0$
  \While{$d < H-1$}
      \State $mid \gets L + \lfloor (R-L)/2 \rfloor$
      \If{$r_q \le mid$}
          \State $R \gets mid$
      \ElsIf{$l_q \ge mid+1$}
          \State $L \gets mid+1$
      \Else
          \State \textbf{break}
      \EndIf
      \State $d \gets d+1$
  \EndWhile

  \State $L_{start} \gets d$
  \State $L_{end} \gets L_{start}$

  \For{$h$ from $L_{start}+1$ to $H-1$}
      \If{\textsc{BoundaryRefinementInvalid}$(l_q,r_q,h,N)$}
          \State \textbf{break}
      \Else
          \State $L_{end} \gets h$
      \EndIf
  \EndFor

  \State \Return $(L_{start},L_{end})$
\end{algorithmic}
\end{algorithm}

\textsc{HotspotLocalization} identifies, for each query, a small query-relevant portion of the ENCs from which ENs are selected under the range constraint. Since a vertex's ENCs are organized across hierarchical layers, many layers are misaligned with the query and unlikely to yield valid or effective ENs. FROG therefore computes a query-specific hotspot once at the start of the search and reuses it for all expanded vertices. When expanding a vertex, FROG selects ENs only from this hotspot, reducing global-memory accesses and subsequent examination overhead.

Using the attribute-to-rank mapping defined in \autoref{sec:index_structure}, a query range becomes a rank interval $[l_q,r_q]$. Each vertex belongs to one segment per layer, with deeper layers covering finer rank intervals. The query hotspot can therefore be represented as a contiguous layer interval $[L_{start},L_{end}]$.

As shown in Algorithm~\autoref{alg:hotspot_localization}, the rules for determining $[L_{start}, L_{end}]$ are as follows.

{\underline{Rule 1:} $L_{start}$ is defined as the shallowest layer at which the query range $[l_q,r_q]$ contains at least one segment boundary. Layers above $L_{start}$ contain segments significantly larger than the query range, so their ENCs are poorly aligned with the range constraint and unlikely to yield useful ENs. This rule makes the segment granularity comparable to the query range.}

\underline{Rule 2:} Starting from $L_{start}$, FROG determines the ending layer $L_{end}$ by requiring that a consecutive sequence of layers satisfy a validity condition. Let $b_l$ and $b_r$ denote the leftmost and rightmost segment boundaries that fall inside $[l_q,r_q]$ at layer $h$, respectively. A layer $h$ is considered valid only if
\begin{equation}
\label{eq:endcondition}
    (b_l-l_q) + (r_q-b_r) \geq 2^{-h}(r_q-l_q)
\end{equation}
FROG scans layers in increasing depth starting from $L_{start}$ and includes each layer in the hotspot as long as this condition holds. Once the condition is violated at some layer, the scan stops, and the last consecutive valid layer is regarded as $L_{end}$.
The left-hand side of \autoref{eq:endcondition} measures the query's total overlap with the boundary segments at layer $h$, while the right-hand side is a conservative threshold, derived from the worst-case boundary segments, that scales with both query length and segment granularity. A violation indicates that the query overlaps only a small portion of the boundary segments and that their ENCs are unlikely to contribute effective ENs; scanning deeper layers would therefore add global-memory accesses and examination cost with little benefit. Consequently, shallow layers typically suffice for large ranges, whereas small ranges can activate deeper, better-aligned layers. Together, the two rules restrict EN identification to a compact, query-specific hotspot, reducing both ENC loading and subsequent condition checks.

\subsection{Expansion Neighbor Identification}
\label{sec:ENIdentification}

The ENCs in the hotspot are further examined through only a few simple logical checks: (1) whether the maximum number of ENs has already been reached; (2) whether the ENC lies within the query range; and (3) whether its distance has already been computed. 

Each thread examines one ENC in parallel, but, as discussed in \autoref{sec:preli_gpu}, the resulting multi-condition branching can cause warp divergence; FROG therefore uses control-flow compaction followed by lazy checking to alleviate these stalls.

\begin{algorithm}[t]
\small
\caption{\textsc{ENIdentification}$(ENC,u,l_q,r_q,L_{start},L_{end},B)$}
\label{alg:en_identification}
\begin{algorithmic}[1]
\Require ENC of $u$, query range $[l_q,r_q]$,
         hotspot layers $[L_{start},L_{end}]$,
         expansion budget $B$
\Ensure expansion neighbors $EN(u)$

\State $EN(u) \gets \emptyset$
\State $lane \gets$ lane id in the warp

\For{$h$ from $L_{start}$ to $L_{end}$}
    \For{each warp-sized chunk $C \subset ENC_h(u)$}
        \State Each lane loads one candidate $v \in C$
        \State $pred \gets (v \neq \bot) \land (l_q \le v \le r_q)$

        \State $M \gets$ \texttt{\_\_ballot\_sync}$(pred)$
        \State $n \gets$ \texttt{\_\_popc}$(M)$
        \State $rank \gets$ \texttt{\_\_popc}$(M \land ((1 \ll lane)-1))$
    
        \State $base \gets |EN(u)|$
        \State $a \gets \min(n, B-base)$
      
        \If{$pred$ and $rank<a$}
            \State write $v$ to $EN(u)[base + rank]$
        \EndIf
        \State The first lane sets $|EN(u)| \gets base+a$
    
        \If{$|EN(u)| = B$}
            \State \Return $EN(u)$
        \EndIf
    \EndFor
\EndFor
\State \Return $EN(u)$
\end{algorithmic}
\end{algorithm}

\subsubsection{Control-flow compaction}

Control-flow compaction is a parallel data-reorganization technique. Given an input sequence and a Boolean predicate defined over its elements, the operation selects all elements satisfying the predicate and packs them into a dense output sequence. Before applying control-flow compaction, consecutive lanes in a warp are mapped to consecutive positions in the candidate ENC array. Since a large fraction of ENCs fail the filtering conditions, only a subset of lanes produce valid results. Compaction assigns these valid candidates consecutive positions in the EN array even when they originate from scattered positions in the ENC array. Although the input ENC chunks must still be scanned, the resulting dense EN array allows the subsequent hash-checking and distance-computation stages to use fewer, more fully utilized warps, leaving idle lanes only at the tail of the last partially filled warp.

Algorithm~\autoref{alg:en_identification} illustrates the warp-level realization of this idea. Although the pseudocode is written at lane (thread) granularity, it should not be interpreted as a sequence of isolated per-thread instructions. Instead, one warp cooperatively and sequentially scans the hotspot in warp-sized chunks, with each lane processing one ENC in the current chunk. Because only one warp constructs $EN(u)$, no atomic reservation is required.
The algorithm starts by initializing $EN(u)$ and scanning the hotspot layers in warp-sized chunks (Lines~1--5). For each chunk, each lane loads one candidate $v$ and evaluates the lightweight validity and range predicates (Line~6). The warp then performs compaction: \textit{\_\_ballot\_sync} builds a mask of valid lanes, and \textit{\_\_popc} obtains both the number of valid candidates and each lane's compacted rank (Lines~7--9). These ranks map scattered valid candidates to consecutive output offsets.
Next, the warp reads the current output size as the base offset and admits only the candidates allowed by the remaining budget (Lines~10--11). A lane writes its candidate if it is valid and its compacted rank is smaller than the admitted count (Lines~12--13). After the writes, the first lane updates the output size (Line~14). The algorithm terminates once the budget $B$ is reached; otherwise, it returns the accumulated $EN(u)$ after all relevant hotspot layers have been processed (Lines~15--17).

\subsubsection{Lazy Hash Check}
Because logical conjunctions exhibit short-circuit behavior, expensive predicates are typically placed later. Accordingly, we originally placed the hash check after the other conditions in line 6. This check determines whether a candidate's distance to the query vector has already been computed and thus avoids redundant computation, but it is considerably more expensive than arithmetic or range tests because it involves hash computation, bucket lookup, and possibly collision resolution. On GPUs, directly applying this “expensive predicates last” strategy remains inefficient: as long as one thread in a warp performs a hash query, threads already pruned by lightweight predicates must wait, and collisions can amplify this delay through additional probes.

To address this issue, FROG postpones the hash check from the candidate-filtering stage to immediately before distance computation. During candidate filtering, only lightweight predicates are used for preliminary pruning, allowing already evaluated nodes to temporarily enter the EN array. The hash table is queried only when a candidate node is placed in the EN array and is about to undergo distance computation.

The key advantage is that candidates requiring high-latency hash checks are concentrated into a small number of warps. Such candidates are sparse in the ENC array, where many threads pruned by earlier conditions would otherwise remain blocked; after compaction into the EN array, every active thread performs the check. The expensive operation is thus moved to a stage with lower concurrency and more balanced inter-thread workloads, reducing both the number of threads subject to ineffective waiting and the waiting duration. Although this strategy may admit a small number of duplicate candidates and require temporary storage, these costs are negligible relative to the reduction in warp stalls; relocating duplicate checking therefore improves GPU execution efficiency while preserving correctness.

\begin{figure*}[t]
  \centering
  \includegraphics[width=\textwidth]{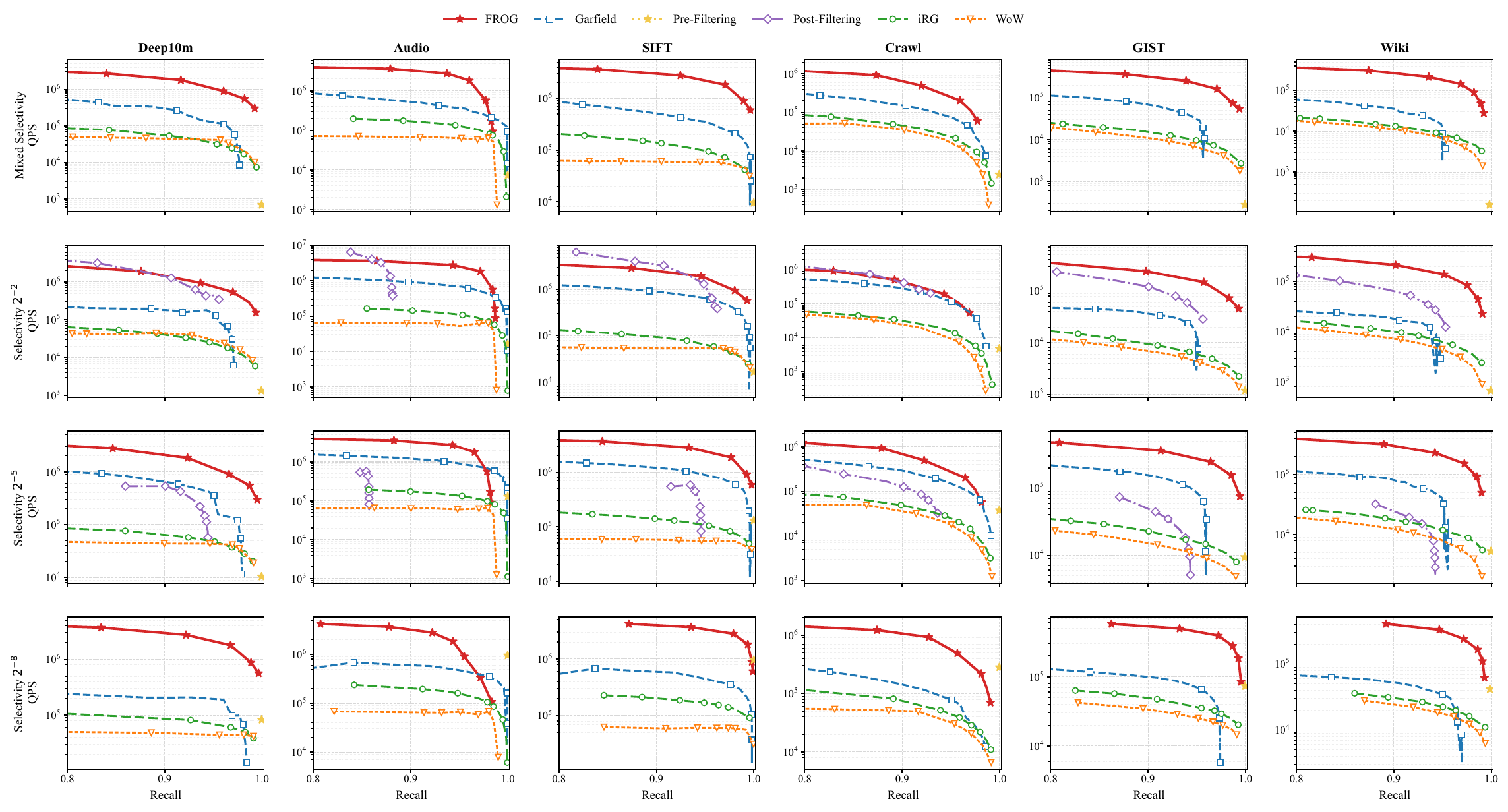}
  \caption{Search throughput versus recall across all datasets under varying selectivity settings.}
  \label{fig:main-search}
\end{figure*}

\section{Experiments}
\label{sec:experiments}
\subsection{Experimental Setting}
We examine the performance of FROG and existing methods on a single attribute. Each experiment is repeated 30 times, and the average is reported.

\subsubsection{Environment}
The CPU-based methods were executed on a machine equipped with dual Intel Xeon Gold 6238 CPUs (44 cores and 88 threads in total). All CPU implementations were parallelized using OpenMP~\cite{openmp50}, with the maximum number of threads set to 88 for index construction and 44 for search. GPU-based methods were conducted on a single NVIDIA RTX 5090, where one CPU thread was dedicated to launching device commands.

\subsubsection{Datasets}
Table~\ref{tab:datasets} summarizes the datasets used in our experiments. 
They cover a wide range of scales $n$ and dimensionalities $d$, with source formats including image, audio, text, and multimodal data.
For real-world datasets, Audio uses the number of likes as the attribute, while Wiki uses text length. 
For synthetic datasets, following prior work~\cite{SeRF, DSG, iRG, WoW}, we randomly shuffle the base vectors and assign their IDs as attribute values.

We evaluate four different query-range settings: a mixed setting, where selectivities $2^{-i}$ with $i \in [0,9]$ are each sampled with equal probability, and three fixed-selectivity settings of $2^{-2}$, $2^{-5}$, and $2^{-8}$, representing large, medium, and small selectivities, respectively.
The query ranges and ground truth are generated using the program provided by~\cite{iRG}.

\subsubsection{Metrics}
Accuracy is measured by Recall@10, which indicates how many of the top-10 ground-truth neighbors are successfully retrieved. Throughput is measured in queries per second (QPS).

To ensure a fair comparison among different methods, the scope of time measurement is carefully defined. 
(1) The time measurement for index construction starts from the initial state where all data reside on disk and ends when the system is ready to serve queries. Therefore, for GPU-based methods, we report the total time including both data transfer and index construction.
(2) Since deployed systems keep the index and database resident in GPU memory, cost at query time includes only the data (query vectors and filter ranges) transfer and execution.

\subsubsection{Competitors and Parameters}
For competitive CPU baselines, we selected \underline{iRG~\cite{iRG}} and the recently published \underline{WoW~\cite{WoW}}. Since these methods do not natively support multi-threaded query processing, we modified their implementations using OpenMP~\cite{openmp50}.
On the GPU side, we implemented \underline{Pre-Filtering} and \underline{Post-Filtering} strategies based on cuVS~\cite{cuvs, cuvs_filtering_docs}. In post-filtering, the top-$k$ neighbors that satisfy the filter are selected from the CAGRA~\cite{CAGRA} search results, where parameters are adaptively configured.
We also include \underline{Garfield~\cite{Garfield}}, the only known GPU-tailored RFANNS method.
Given the strong support of GPUs for half-precision floating-point operations, all GPU programs use \textit{float16} for distance computations, while the accumulated distance values are stored as \textit{float32}.

For the general parameters, we slightly adjusted the settings of prior work to better align with GPU memory practices. Specifically, for datasets with dimensionality $\leq 300$, we set $m=16$ and $\mathrm{ef_c}=128$. For datasets with dimensionality $> 300$, we set $m=32$ and $\mathrm{ef_c}=256$.
The method-specific parameters of each baseline are configured according to their recommended settings. 

For FROG-specific parameters, we did not conduct fine-grained tuning; instead, we adopted a set of empirically validated default configurations. We set the number of omitted deepest layers to $n_{\mathrm{inv}} = 7$ for datasets with dimensionality $\leq 300$ and $n_{\mathrm{inv}} = 6$ for datasets with dimensionality $> 300$. Across all experiments, we set the fusion distance parameters to $\beta = 0.2$ and $\gamma = 0.5$, the number of entry points to $epn=16$, the expansion budget to $B=16$, and the \textit{patience} to 30.

\begin{table}[h]
    \centering
    \caption{Summary of datasets used in the experiments.}
    \label{tab:datasets}
    \begin{tabular}{cccc}
    \toprule
    \textbf{Dataset} & \textbf{Scale} & \textbf{Dimension} & \textbf{Source format} \\
    \midrule
    Deep10m~\cite{deep}   & 10,000,000 & 96   & Image \\
    Audio~\cite{youtube}  & 1,000,000  & 128  & Audio \\
    SIFT~\cite{siftgist}  & 1,000,000  & 128  & Image \\
    Crawl~\cite{crawl}    & 1,989,995  & 300  & Text \\
    GIST~\cite{siftgist}  & 1,000,000  & 960  & Image \\
    Wiki~\cite{wiki}      & 1,000,000  & 2048 & Image \& Text \\
    \bottomrule
    \end{tabular}
\end{table}

\subsection{Experimental Results}
\subsubsection{Search Performance}
\autoref{fig:main-search} shows the experimental results of search performance. 
Several facts can be inferred:

(1) FROG achieves the best performance in almost all settings. Using QPS interpolated at Recall = 0.9, FROG achieves a 14.7$\times$--37.7$\times$ speedup over CPU methods and a 4.5$\times$--7.6$\times$ speedup over the current GPU state of the art under mixed selectivity across the six datasets.

(2) Methods based on general-purpose GPU indexes have inherent limitations.
Pre-filtering achieves acceptable performance when selectivity is low, but breaks down under high-selectivity conditions.
Post-filtering struggles to achieve sufficiently high recall in many scenarios. Although it can sometimes outperform all other methods on simple datasets, under extremely high selectivity, or when the recall target is low, this advantage relies on prior knowledge of the selectivity. In mixed-selectivity settings, post-filtering also fails to provide sufficient concurrency, as it requires query-specific parameter tuning for each query.

This indicates that relying solely on GPU computational power is insufficient; a carefully designed index is necessary.

(3) Across different selectivities, FROG exhibits substantially smaller performance fluctuations than other methods and does not exhibit any failure cases, demonstrating strong robustness.

\subsubsection{Construction Performance}
\begin{figure}[t]
  \centering
  \includegraphics[width=\linewidth]{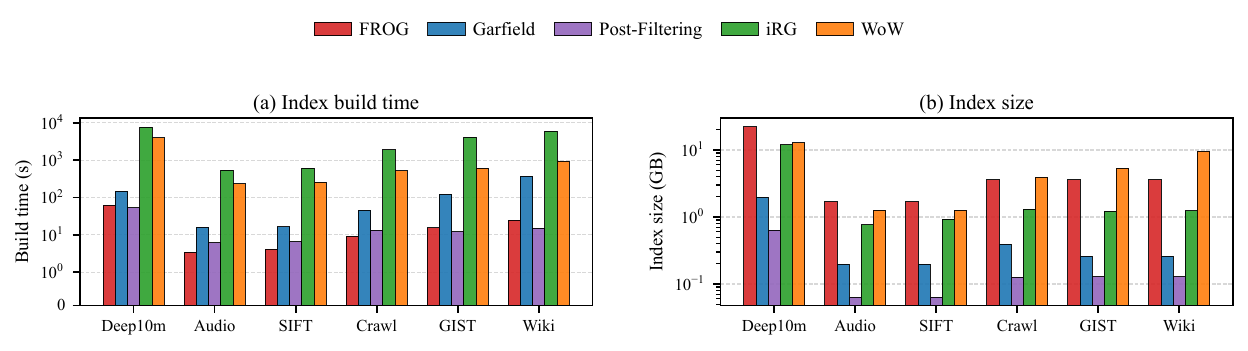}
  \caption{Index construction overhead across all datasets.}
  \label{fig:main-build}
\end{figure}

The construction runtime and index size of the indexes used in the query performance experiments are shown in \autoref{fig:main-build}. Methods that do not build specialized indexes are excluded. Across the datasets, FROG accelerates index construction by 37.1$\times$--68.4$\times$ over WoW and by 2.4$\times$--14.8$\times$ over the current GPU state of the art. Despite its memory-alignment requirements, FROG maintains acceptable space overhead by removing redundant layers.

\begin{figure}[t]
  \centering
  \includegraphics[width=\linewidth]{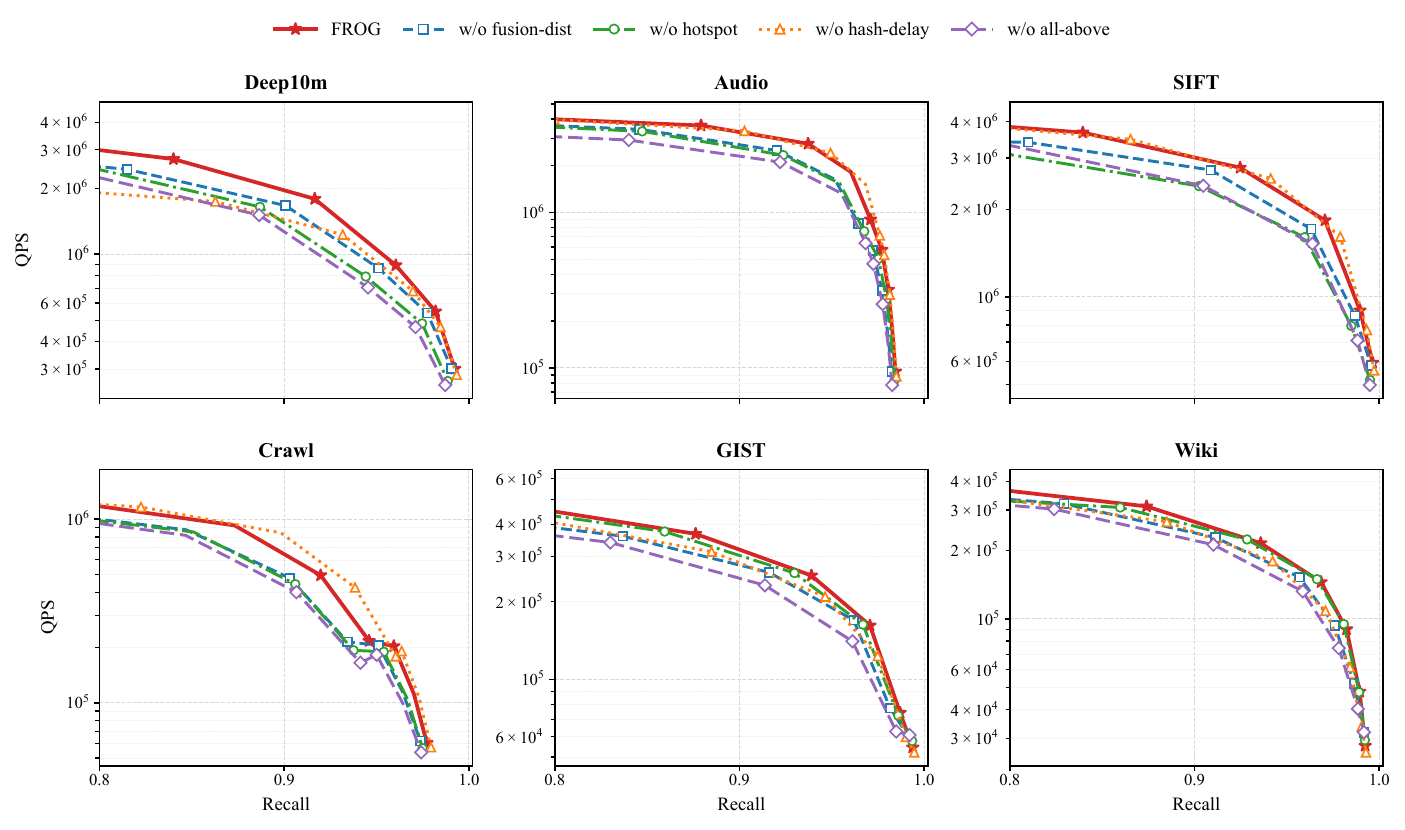}
  \caption{Ablation study under mixed selectivity. The recall--QPS curves compare
  full FROG with variants that remove fusion distance, hotspot handling, hash
  delay, or all three components.}
  \label{fig:ablation-search}
\end{figure}

\begin{figure}[t]
  \centering
  \includegraphics[width=\linewidth]{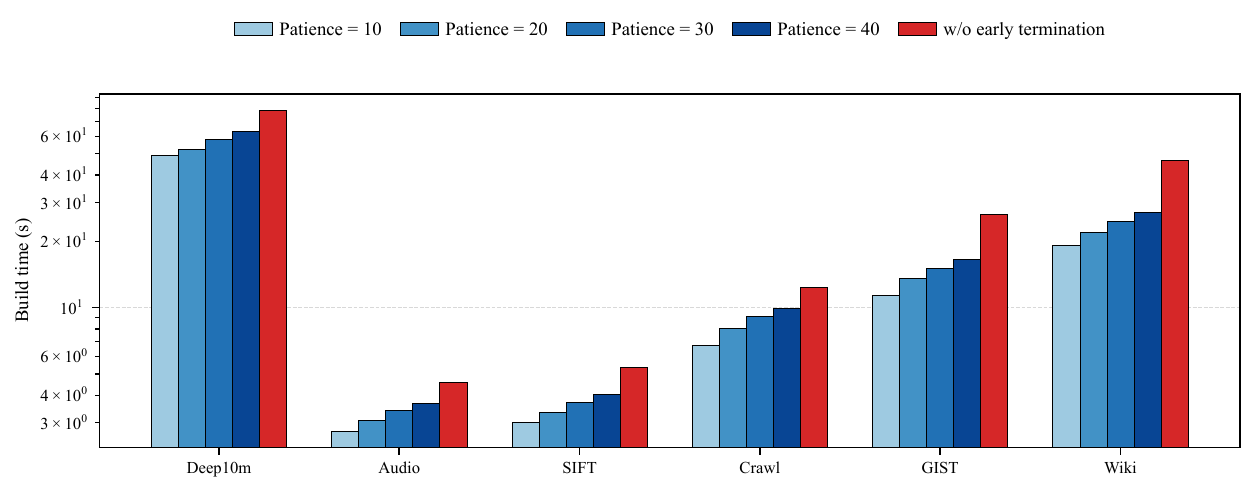}
  \caption{Effect of the early-termination patience parameter on index build
  time across all datasets.}
  \label{fig:patience-build-time}
\end{figure}

\begin{figure}[t]
  \centering
  \includegraphics[width=\linewidth]{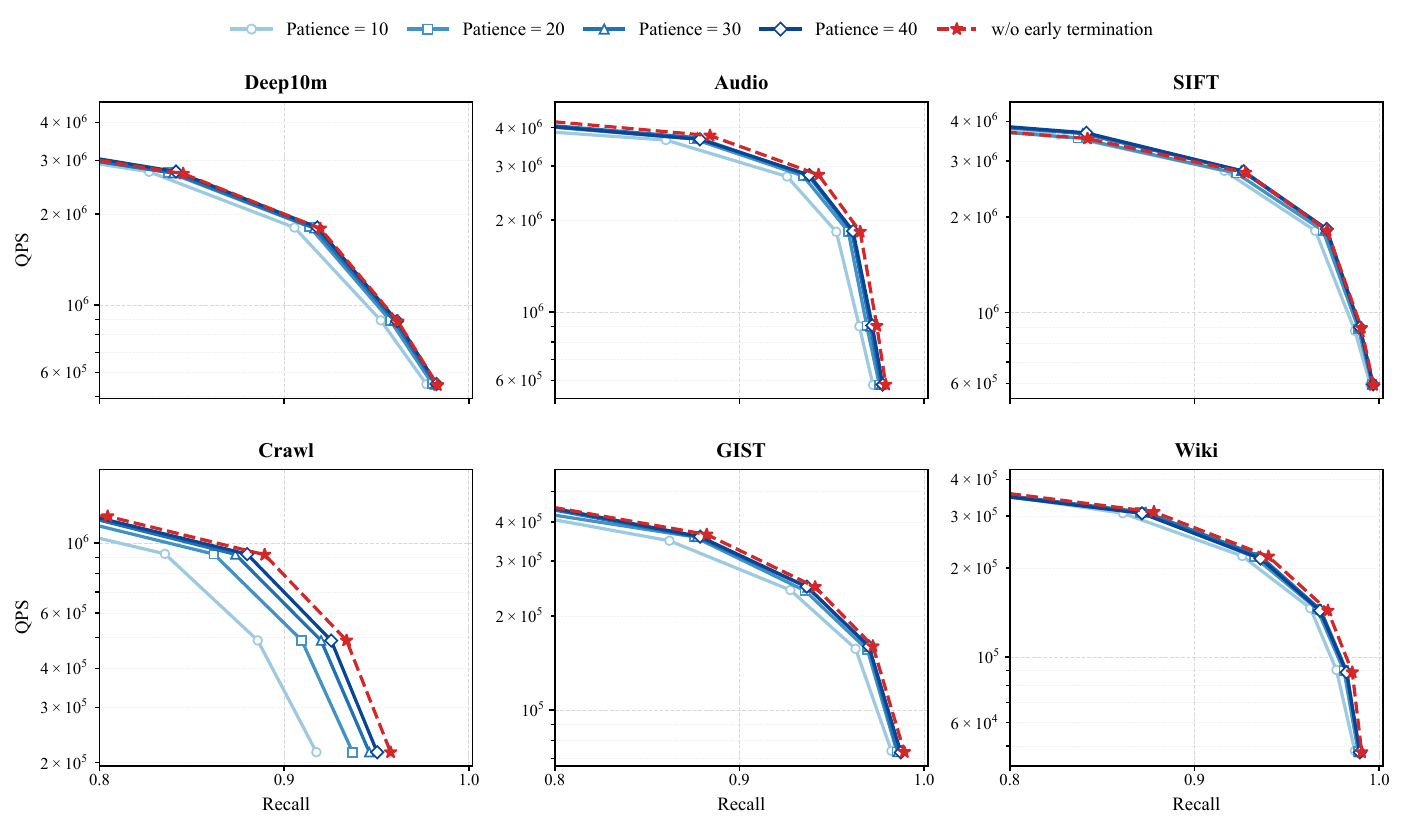}
  \caption{Effect of the early-termination patience parameter on search
  throughput and recall across all datasets.}
  \label{fig:patience-search}
\end{figure}

\subsubsection{Ablation Study}
We conducted an ablation study to evaluate three key optimizations: fusion distance, hotspot localization, and lazy hash checking. \autoref{fig:ablation-search} illustrates the performance impact of enabling or disabling each optimization.

Overall, disabling these components weakens the recall--throughput trade-off, while removing all of them causes the most pronounced degradation. This confirms that FROG's ENC-selection criterion and GPU-oriented query optimizations complement each other in improving search efficiency.

\subsubsection{Sensitivity Analysis of \textit{patience}}
\autoref{fig:patience-build-time} and \autoref{fig:patience-search} show that reducing \textit{patience} substantially accelerates index construction while having only a limited impact on query performance. However, Crawl is more difficult and converges more slowly, making aggressive early termination more likely to discard useful neighbors. We therefore set \textit{patience} to 30, which provides a good balance between construction efficiency and search quality.

\subsubsection{Sensitivity Analysis of $\beta$ and $\gamma$}
\begin{figure}[t]
  \centering
  \setlength{\tabcolsep}{2pt}
  \captionsetup[subfigure]{font=small, skip=2pt}

  \begin{subfigure}[t]{0.2\textwidth}
    \centering
    \includegraphics[width=\linewidth]{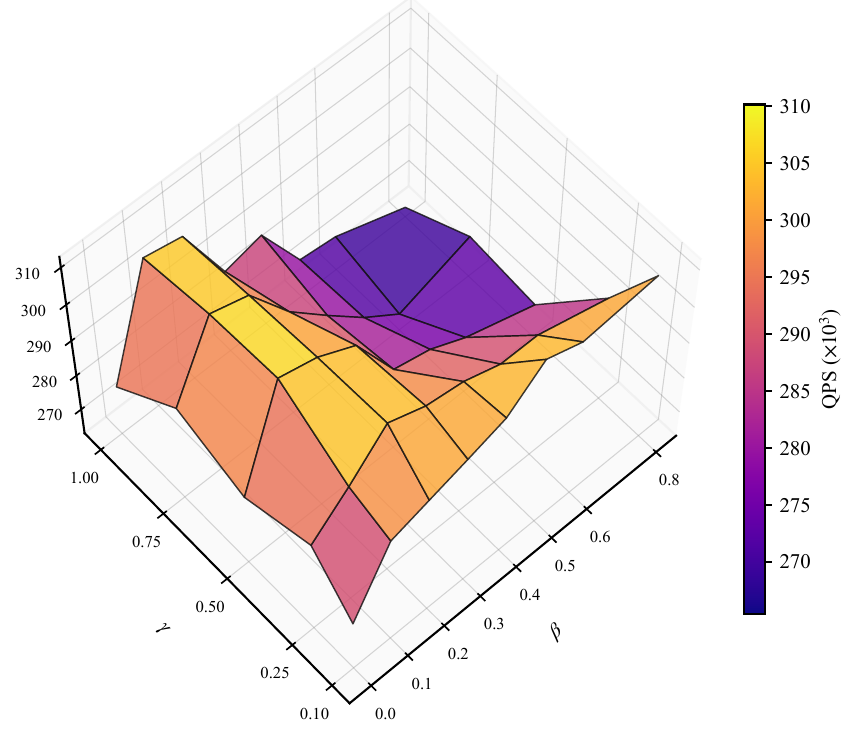}
    \caption{GIST, recall $=0.90$.}
    \label{fig:fusion-dist-gist-090}
  \end{subfigure}\hfill
  \begin{subfigure}[t]{0.2\textwidth}
    \centering
    \includegraphics[width=\linewidth]{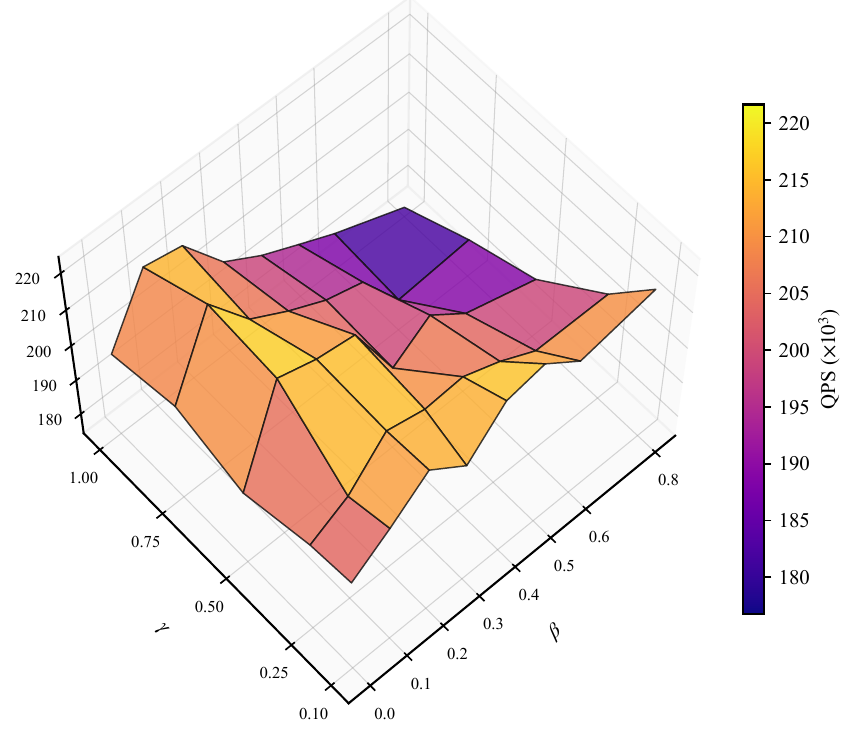}
    \caption{GIST, recall $=0.95$.}
    \label{fig:fusion-dist-gist-095}
  \end{subfigure}

  \medskip

  \begin{subfigure}[t]{0.2\textwidth}
    \centering
    \includegraphics[width=\linewidth]{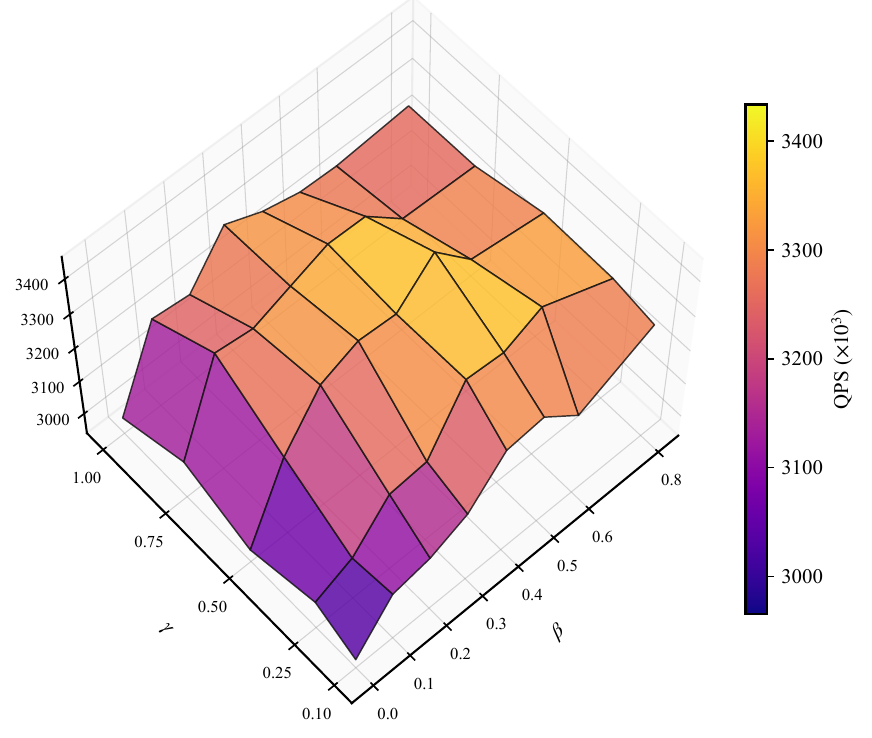}
    \caption{Audio, recall $=0.90$.}
    \label{fig:fusion-dist-audio-090}
  \end{subfigure}\hfill
  \begin{subfigure}[t]{0.2\textwidth}
    \centering
    \includegraphics[width=\linewidth]{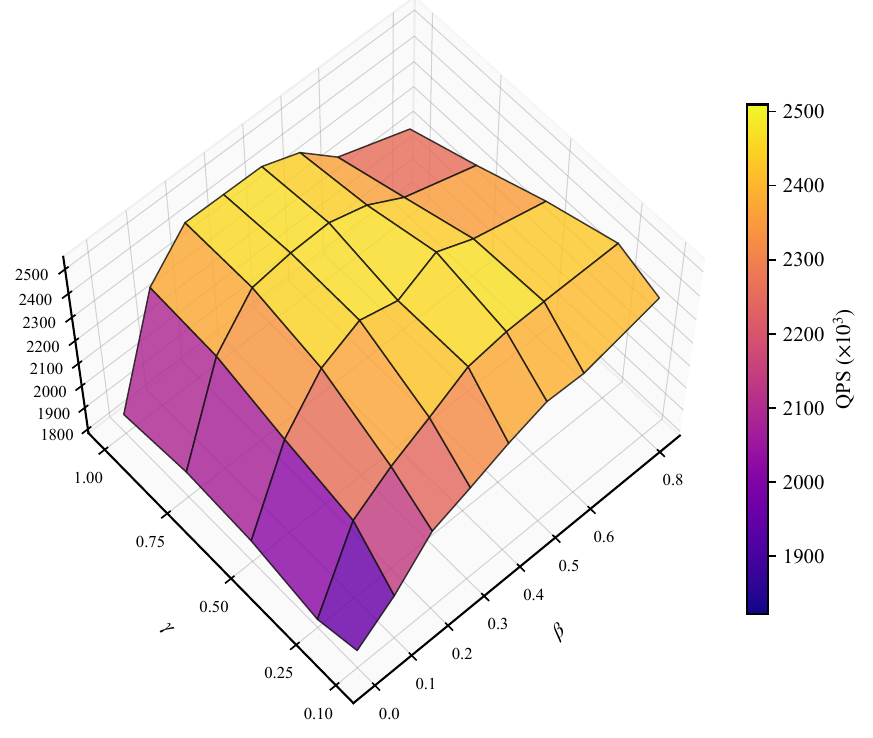}
    \caption{Audio, recall $=0.95$.}
    \label{fig:fusion-dist-audio-095}
  \end{subfigure}

  \caption{Sensitivity of FROG's search throughput to the fusion-distance
  parameters $\beta$ and $\gamma$ at two target recall levels.}
  \label{fig:fusion-dist-study}
\end{figure}

\autoref{fig:fusion-dist-study} shows how query performance on the Audio and GIST datasets varies with different values of the fusion-distance parameters $\beta$ and $\gamma$. The results indicate that performance is more sensitive to $\beta$ than to $\gamma$: a moderate value of $\gamma$ is usually sufficient, whereas $\beta$ should be selected according to the data dimensionality. For low-dimensional datasets, performance exhibits a broad plateau, where a wide range of $\beta$ values works well. In contrast, high-dimensional datasets achieve peak performance only along a much narrower ridge.

At each fixed $\gamma$, introducing a moderate positive $\beta$ generally improves QPS over $\beta=0$, whereas an excessively strong fusion bias eventually degrades performance. The optimal $\beta$ also tends to decrease as $\gamma$ increases. These trends are consistent with the theoretical prediction in \autoref{sec:fusion_criterion}: fusion selection initially improves valid-ENC availability, while excessive attribute weighting eventually incurs vector-distance distortion.

This experiment provides guidance for parameter selection. When exact tuning is not required, the empirical settings $\beta = 0.2$ and $\gamma = 0.5$ are sufficiently robust across datasets.

\begin{figure}[t]
  \centering
  \includegraphics[width=\linewidth]{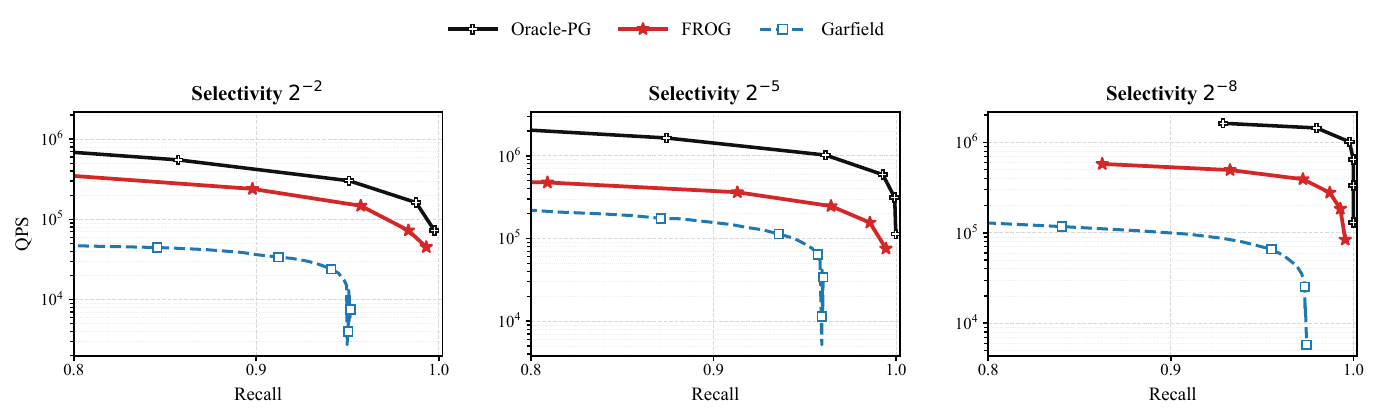}
  \caption{Recall--QPS comparison with the Oracle PG performance on GIST
  at selectivities $2^{-2}$, $2^{-5}$, and $2^{-8}$.}
  \label{fig:ceiling-search}
\end{figure}

\subsubsection{Upper-bound Comparison}
We compare FROG with Oracle PG, which has prior knowledge of the qualifying vectors and searches on a dedicated unfiltered graph built over them, thereby providing an empirical performance ceiling. As shown in Figure~\ref{fig:ceiling-search}, FROG substantially narrows the gap to this performance ceiling compared with existing methods.

\section{Related Work}
\paragraph{Attribute-Filtering ANNS}
According to a recent survey~\cite{AF_survey}, attribute-constrained ANNS can be broadly classified into range-filtering $k$-ANN search (RFANNS), which imposes numerical range constraints, and label-filtering $k$-ANN search (LFANNS), which imposes categorical constraints.
General-purpose systems~\cite{ACORN,faiss,cuvs,Milvus} support both types of constraints, whereas specialized methods are tailored to either LFANNS~\cite{NHQ,fdiskann} or RFANNS~\cite{SeRF,DSG,UNIFY,WST,iRG,WoW,DIGRA}.

\paragraph{GPU-accelerated Vector Search}
GPU-based vector search methods exploit massive parallelism to improve query throughput. Existing graph-based approaches~\cite{SONG,GANNS,GGNN,CAGRA} redesign ANNS indexes and search procedures for efficient GPU execution, while other systems~\cite{BANG,rummy,PilotANN,Tagore} further balance throughput and scalability. For attribute-filtering ANNS on GPUs, VecFlow~\cite{vecflow} targets LFANNS, while Garfield~\cite{Garfield} is designed specifically for RFANNS.

\section{Conclusion}
In this paper, we propose FROG, a globally aware, vertex-centric GPU index for RFANNS. FROG stores range-diverse ENCs for each vertex and identifies query-specific ENs online to approximate the Oracle PG induced by each query range. Its fusion-distance criterion, bottom-up parallel construction, hotspot localization, control-flow compaction, and lazy hash checking jointly make RFANNS search efficient on GPUs.
Across six datasets and multiple selectivities, FROG delivers robust recall-throughput performance. At Recall@10 = 0.9, it improves query throughput by up to 37.7$\times$ over 44-core CPU baselines and by up to 7.6$\times$ over the strongest GPU baseline. It also reduces index construction runtime by up to 68.4$\times$ compared with WoW and by up to 14.8$\times$ compared with the GPU baseline.


\FloatBarrier
\bibliographystyle{ACM-Reference-Format}
\balance
\bibliography{sample}

\end{document}